\documentclass[11pt]{article}

\usepackage[utf8]{inputenc}
\usepackage[T1]{fontenc}
\usepackage[right=1in,left=1in,top=1.1in,bottom=1.1in]{geometry}
\usepackage{amsmath,amssymb,amsthm,mathtools}
\usepackage{enumitem}
\usepackage{setspace}
\usepackage{hyperref}
\hypersetup{
  colorlinks=true,
  citecolor=blue,
  linkcolor=blue,
  urlcolor=blue
}

\usepackage[authoryear,round]{natbib}

\newtheorem{theorem}{Theorem}[section]
\newtheorem{lemma}[theorem]{Lemma}
\newtheorem{prop}[theorem]{Proposition}
\newtheorem{assumption}{Assumption}[section]
\newtheorem{definition}[theorem]{Definition}
\theoremstyle{definition}
\newtheorem{example}{Example}[section]
\newtheorem*{example*}{Example}

\DeclareMathOperator{\Var}{Var}
\DeclareMathOperator{\Cov}{Cov}
\newcommand{\var}[1]{\Var\left[#1\right]}

\title{Nonparametric Subgraph Regression}

\author{Olivia Ding\footnote{
Email: \href{mailto:olivia.ding@yale.edu}{olivia.ding@yale.edu};
I am grateful to Yuichi Kitamura, Xiaohong Chen, Tim Christensen,
Ed Vytlacil, P. Aronow, Don Andrews, and Max Cytrynbaum for their
invaluable advice and encouragement. I thank ChatGPT and Claude for research assistance in literature review, proof checking, and code generation.}\\
Yale University}

\date{This version: \today\\
(First version: May 17, 2024)}

\begin{document}
\onehalfspacing
\maketitle
\vspace{-2.5em}
\begin{center}
\href{https://drive.google.com/file/d/1cb_fadqqXOdt5Ihk-kzIefNFoEtAl6Cb/view?usp=sharing}{[Click here for the latest version]}
\end{center}

\begin{abstract}
Estimation and inference on a single large network are challenging due
to dependence among observations. This paper studies nonparametric
regression for outcomes defined on groups of $p$ agents, such as the
indicator that three agents in a network form a triangle. Outcomes on
groups that share an agent may be dependent, while outcomes on disjoint
groups are independent. I call this problem subgraph regression; it
nests dyadic regression as the special case $p=2$. Its leading
application concerns
subgraphs of a network, but the methods apply to any outcome defined on
groups of agents. I establish a variance
bound, uniform convergence rates in probability, and pointwise
asymptotic normality for the subgraph analogue of the Nadaraya-Watson
kernel regression estimator. The setting differs from regression with
i.i.d.\ data in two ways: (i) the number of observations, of order
$N^p$, overstates the effective sample size, which is $N$, and (ii) the
rate of convergence depends on the dimension of the covariates of a
single agent rather than of all $p$ agents, which mitigates the
curse of dimensionality.
\end{abstract}


\section{Introduction}\label{sec:intro}

Many economic outcomes belong to groups of agents rather than to
individuals, and the groups come in different sizes. Trade flows and
military alliances link pairs or larger coalitions of countries
\citep{poast_misusing_2010}; risk-sharing arrangements bind pairs or
small groups of households; and a transitive friendship involves three
students at once. A recurring empirical question is how the
distribution of such an outcome varies with the characteristics of the
agents involved. For instance, how
does the probability that three students are all friends with one
another depend on their
ages, incomes, and ethnicities? Subgraph frequencies of this kind are
central to the analysis of network formation, where clustering and
transitivity are leading summaries of strategic and homophilous linking
\citep{jackson_social_2008,mele_structural_2017,sheng_structural_2020,
graham_network_2020}. This paper develops nonparametric regression
methods for such outcomes when they are observed on a single large
network.

The setting is as follows. A researcher samples \(N\) agents from a large
population and observes a vector of covariates \(X_i\in\mathbb R^{d_X}\)
for each agent. For every ordered \(p\)-tuple of distinct agents
\(\mathbf i_p=(i_1,\dots,i_p)\), the researcher also observes a \emph{polyadic} outcome \(Y_{\mathbf i_p}\). The object of interest is the regression function
\[
m(w)=E\bigl[Y_{\mathbf i_p}\mid X_{\mathbf i_p}=w\bigr],
\]
for \(w\in\mathbb R^{pd_X}\), where \(X_{\mathbf i_p}=(X_{i_1}',\dots,X_{i_p}')'\in\mathbb R^{pd_X}\) stacks the
covariates of the \(p\) agents. Other than smoothness, \(m\) is left
unrestricted. The running example is the conditional subgraph
probability. For a fixed pattern \(S\) of links among \(p\) agents,
such as a triangle, \(Y_{\mathbf i_p}\) is the indicator that the links
among the agents in \(\mathbf i_p\) form \(S\), and \(m(w)\) is
the probability of this pattern given the agents' covariates. I use the
term \emph{subgraph regression} because this is the leading
application. Nothing in the theory, however, requires the outcome to be
a function of the network or a binary variable. The results apply to any polyadic outcome
that satisfies the moment conditions of Section~\ref{sec:framework},
including the trade, alliance, and risk-sharing outcomes above.

In this setting, outcomes on disjoint tuples are independent, but those on tuples that
share an agent are generally dependent through the characteristics of
the agents they share. In the terminology of
\citet{davezies_empirical_2021}, this structure is referred to as
\emph{polyadic dependence}. It arises under the exchangeable random
graph models that underlie much of the econometrics of networks
\citep{aldous_representations_1981,lovasz_limits_2006,
graham_network_2020,kitamura_estimating_2024}. When \(p=2\), it
reduces to the dyadic dependence studied by
\citet{aronow_cluster-robust_2015} and \citet{graham_minimax_2021}.

Polyadic dependence complicates estimation and inference in two ways.
First, the \(N!/(N-p)!\) observed outcomes contain far less information
than an independent sample of the same size, because every agent
appears in \(O(N^{p-1})\) tuples. Second, standard tools for
nonparametric regression with dependent data, which rely on mixing or on
a temporal or spatial ordering, do not apply. For \(p=2\),
\citet{graham_minimax_2021} show that dyadic dependence changes both the
effective sample size and the effective dimension of the estimation
problem. To the best of my knowledge, no corresponding results are available for
groups of three or more agents.

I study the subgraph analogue of the Nadaraya-Watson kernel regression
estimator and make four contributions. First, I show that the variance
of the kernel numerator is of order \((Nh_N^{d_X})^{-1}\), where \(h_N\)
is the bandwidth. The bound follows from counting overlapping tuples and
from localized moment conditions. Second, I establish that the kernel
numerator converges uniformly in probability at the rate
\((\ln N/(Nh_N^{d_X}))^{1/2}\) over a region that expands at a
polynomial rate. The proof combines the covering
argument of \citet{hansen_uniform_2008} with a truncation argument for
heavy-tailed outcomes. Bernstein's inequality controls the H\'ajek
projection, the exponential inequalities of \citet{arcones_limit_1993}
control the degenerate components of the Hoeffding decomposition, and
a bounded-differences inequality controls the component driven by the
link shocks. Third, I combine these results with a bias calculation for higher-order kernels to obtain
the uniform convergence rate of the regression estimator.
Fourth, I establish pointwise asymptotic normality of the estimator at
the rate \((Nh_N^{d_X})^{1/2}\). The asymptotic variance is determined
by the first Hoeffding projections of the numerator and the denominator.
To the best of my knowledge, no such result is available even for dyadic
regression (\(p=2\)). \citet{graham_minimax_2021} establish minimax
lower bounds and convergence rates for dyadic regression but do not
derive a limiting distribution. The distributional results of
\citet{graham_kernel_2019}, \citet{chiang_empirical_2023}, and
\citet{cattaneo_uniform_2024} concern dyadic density estimation rather
than regression.

The results deliver two lessons, both of which extend those of
\citet{graham_minimax_2021} from dyads to groups of any size. First, the
relevant sample size is the number of agents, \(N\), not the number of
tuples, which is of order \(N^p\). This formalizes, for nonparametric
regression with polyadic data, the intuition that dependence between
overlapping observations makes inference less precise than a naive count of observations would
suggest \citep{aronow_cluster-robust_2015}. Second, the convergence rate
depends on \(d_X\), the dimension of a single agent's covariates, rather
than on \(pd_X\), the dimension of the argument of \(m\). With \(N\)
i.i.d.\ observations and \(pd_X\) regressors, the optimal uniform rate
would be \((\ln N/N)^{\beta/(2\beta+pd_X)}\) \citep{stone_optimal_1982}.
The rate obtained here, \((\ln N/N)^{\beta/(2\beta+d_X)}\), is instead
that of a \(d_X\)-dimensional problem, whatever the size of the
group. As \citet{graham_minimax_2021} observe for dyads, the reason
is that the H\'ajek projection of the estimator holds one agent's
covariates fixed and averages over those of the other \(p-1\) agents.
The same averaging improves the rate of partial means and marginal
integration estimators \citep{newey_kernel_1994,linton_kernel_1995}
and appears in the analysis of conditional \(U\)-statistics
\citep{stute_conditional_1991}. For \(p=2\), the uniform rate coincides
with the minimax lower bound of \citet{graham_minimax_2021} when the
covariate density is bounded away from zero, so the estimator is
rate-optimal in the dyadic case. Extending their lower bound to
\(p\ge3\) is left for future work.

The analysis faces technical challenges that do not arise in the dyadic
case. When \(p\ge3\), two distinct tuples can share two or more
agents, and hence the link shocks attached to the shared pairs, rather
than a single agent. The kernel numerator is therefore not a \(U\)-statistic in the
agent primitives. I decompose it into a \(U\)-statistic of order
\(p\) in the agent primitives and a remainder that has mean zero
conditional on those primitives. Given the agent primitives, the
remainder depends only on the independent link shocks, so I bound it
with McDiarmid's bounded-differences inequality and a count of the
agents whose covariates fall in the support of the kernel. The
\(U\)-statistic component has \(p-1\) degenerate Hoeffding projections,
each with its own rate of concentration. These rates determine the
admissible range of truncation thresholds and hence the bandwidth
conditions for the uniform results. For pointwise inference, I treat the
numerator and the random denominator jointly. I obtain the limiting
distribution of the ratio from a H\'ajek reduction, a
Lindeberg--Feller central limit theorem for the leading projection, and
the delta method.

This paper contributes to three literatures. The first is the theory
of nonparametric estimation. With independent observations, it includes
the optimal rates of \citet{stone_optimal_1980,stone_optimal_1982} and
the textbook treatment in \citet{tsybakov_introduction_2008}. With
weakly dependent observations,
it includes the uniform convergence rates of \citet{hansen_uniform_2008}
for kernel estimators and of \citet{chen_optimal_2015} for series
estimators, and the sup-norm rates and uniform inference of
\citet{chen_optimal_2018} for nonparametric instrumental variables. The
second studies
estimation and inference with dyadic and exchangeable array data.
Examples include dyadic density estimation
\citep{graham_kernel_2019,chiang_empirical_2023,cattaneo_uniform_2024},
dyadic regression \citep{graham_minimax_2021}, empirical process theory
for exchangeable arrays \citep{davezies_empirical_2021}, the bootstrap
under multiway dependence \citep{menzel_bootstrap_2021}, and double
machine learning with multiway clustering \citep{chiang_multiway_2019}.
I extend the nonparametric regression results in this literature from
dyads to groups of any fixed size. The third uses subgraph
frequencies to learn about network formation. Examples include the
method of moments based on subgraph counts \citep{bickel_method_2011}
and structural models of network formation identified from subnetworks
\citep{sheng_structural_2020}. The conditional subgraph probabilities
studied here are nonparametric, covariate-dependent analogues of these
frequencies.

The rest of the paper is organized as follows.
Section~\ref{sec:framework} introduces the model and the subgraph
Nadaraya-Watson estimator. Section~\ref{sec:rates} establishes the
variance bound and the uniform convergence rates of the kernel numerator
and the regression estimator. Section~\ref{sec:pointwise} establishes
pointwise asymptotic normality. Section~\ref{sec:conclusion} concludes.
All proofs are in the Appendix.

\section{Nadaraya-Watson estimator for subgraph regression}
\label{sec:framework}

Let $i\in[N]:=\{1,2,\dots,N\}$ index the random sample of $N$ agents. Consider a scalar outcome
  \(Y_{\mathbf i_p}\) for every ordered \(p\)-tuple
  \(\mathbf i_p=(i_1,\dots,i_p)\) of distinct agents. Also, each agent \(i\) has primitives \(\Xi_i:=(X_i',U_i)'\) consisting
of the observed covariates \(X_i\in\mathbb R^{d_X}\) and a latent heterogeneity \(U_i\in\mathbb R\), both of which are continuously distributed and i.i.d. across agents. The covariate vector \(X_i\) has a density \(f\), and I write \(X_{\mathbf i_p}:=(X_{i_1}',\dots,X_{i_p}')'\in\mathbb R^{d_W}\),
  with \(d_W:=pd_X\), for the stacked covariates of its agents. Without loss of generality, I normalize \(U_i\) to be uniformly distributed on \([0,1]\). Moreover, I assume \(U_i\) is independent of \(X_i\).

  Let \(\mathcal P_{p,N}\) be the set of ordered \(p\)-tuples of distinct agents, with
  cardinality \(M_N:=|\mathcal P_{p,N}|=N!/(N-p)!\). For
  each \(\mathbf i_p\in\mathcal P_{p,N}\),
I assume $Y_{\mathbf i_p}$ is generated by the model
\begin{equation}\label{eq:polyadic-outcome}
Y_{\mathbf i_p}
=H\!\left(\Xi_{i_1},\ldots,\Xi_{i_p},
          (V_{\{i_a:\,a\in B\}})_{B\subseteq\{1,\dots,p\},\,|B|\ge2}\right),
\end{equation}
where \(H\) is a fixed measurable function that depends on neither
\(\mathbf i_p\) nor \(N\), and $V$ denotes the idiosyncratic shocks, indexed by different levels ranging from dyads with $|B|=2$ to the whole tuple with $|B|=p$.
The shocks are assumed to be i.i.d.\ uniform on \([0,1]\) and independent of all primitives.
This model nests the dyadic regression model of
\citet{graham_minimax_2021}: when \(p=2\), the only subset with
\(|B|\ge2\) is the pair itself, so \eqref{eq:polyadic-outcome}
reduces to \(Y_{ij}=H(\Xi_i,\Xi_j,V_{\{i,j\}})\), with a single
shock at the dyad level. For \(p\ge3\), the model allows shocks at every
intermediate level, from pairs up to the whole tuple.
This model is also related to the Aldous--Hoover representation of a
dissociated, jointly exchangeable array
\citep{aldous_representations_1981,davezies_empirical_2021}\footnote{Here the covariates enter through the agent primitives, and the regression function conditions on them, which neither paper considers.}.

Two features of the model are worth noting. First, tuples with no agent in common
have no shared primitives or shocks, so their outcomes are
independent. Second, because \(H\) is common to all tuples, and the
primitives and shocks are i.i.d., the data are invariant to relabeling
the agents: for every permutation \(\pi\) of \([N]\),
\begin{equation}\label{eq:exchangeable}
\bigl(X_{\pi(i)},\,Y_{\pi(\mathbf i_p)}\bigr)_{i\in[N],\,
\mathbf i_p\in\mathcal P_{p,N}}
\overset{d}{=}
\bigl(X_i,\,Y_{\mathbf i_p}\bigr)_{i\in[N],\,\mathbf i_p\in\mathcal P_{p,N}},
\end{equation}
with \(\pi(\mathbf i_p):=(\pi(i_1),\dots,\pi(i_p))\).

The statistical problem is to estimate the regression function
\[
m(w):=E[Y_{\mathbf i_p}\mid X_{\mathbf i_p}=w]
\]
for $w\in\mathbb R^{d_W}$ when the only restriction on it is that it belongs to a H\"older class of functions. The framework covers any polyadic outcome of the form
\eqref{eq:polyadic-outcome}. I use the conditional subgraph probability in Example~\ref{ex:subgraph} as a running example throughout the paper, and show how the conditions simplify in this more structured case.

\begin{definition}[H\"older smoothness]\label{def:holder}
Let \(\beta,L>0\), and let
\(\ell:=\max\{r\in\mathbb Z:r<\beta\}\).
For a multi-index \(\eta=(\eta_1,\dots,\eta_{d_W})\), write $|\eta|:=\eta_1+\cdots+\eta_{d_W}$ and the derivative
\[
D^\eta
:=
\frac{\partial^{|\eta|}}
{\partial w_1^{\eta_1}\cdots\partial w_{d_W}^{\eta_{d_W}}}.
\]
A function \(a:\mathbb R^{d_W}\to\mathbb R\) belongs to
\(\Sigma(\beta,L)\) if it is \(\ell\) times continuously differentiable
and, for every multi-index \(\eta\) with \(|\eta|=\ell\) and all \(w,w'\in\mathbb R^{d_W}\),
\[
|D^\eta a(w)-D^\eta a(w')|
\le L\|w-w'\|_\infty^{\beta-\ell}.
\]
For an open set \(\mathcal U\subseteq\mathbb R^{d_W}\), a function
belongs to \(\Sigma(\beta,L)\) on \(\mathcal U\) if it is \(\ell\) times
continuously differentiable on \(\mathcal U\) and the inequality holds
for all \(w,w'\in\mathcal U\).
\end{definition}

Let \(K:\mathbb R^{d_W}\to\mathbb R\) be a fixed multivariate kernel and
\(h_N\) a bandwidth sequence with \(h_N\to0\). A common choice is the
product kernel \(K(u)=\prod_{a=1}^pk(u_a)\) for
\(u=(u_1',\dots,u_p')'\), built from a kernel \(k\) on
\(\mathbb R^{d_X}\); for \(d_X=1\), common choices of \(k\) are the
Epanechnikov kernel \(k(t)=\frac34(1-t^2)\mathbf1\{|t|\le1\}\) and the
Gaussian kernel \(k(t)=(2\pi)^{-1/2}e^{-t^2/2}\).

The kernel weight of a tuple
\(\mathbf i_p\in\mathcal P_{p,N}\) at \(w\in\mathbb R^{d_W}\) is
\[
K_{\mathbf i_p}(w)
:=
h_N^{-d_W}K\!\left(\frac{X_{\mathbf i_p}-w}{h_N}\right).
\]
For \(w=(w_1',\dots,w_p')'\), let \(f_W(w):=\prod_{a=1}^pf(w_a)\) be
the density of \(X_{\mathbf i_p}\), and let \(\Psi(w):=m(w)f_W(w)\).
Define the kernel-weighted averages
\[
\hat\Psi_N(w)
:=
\frac1{M_N}
\sum_{\mathbf i_p\in\mathcal P_{p,N}}
K_{\mathbf i_p}(w)Y_{\mathbf i_p},
\qquad
\hat f_N(w)
:=
\frac1{M_N}
\sum_{\mathbf i_p\in\mathcal P_{p,N}}
K_{\mathbf i_p}(w),
\]
which estimate \(\Psi(w)\) and \(f_W(w)\). I estimate \(m(w)\) by their ratio, a subgraph analogue of the Nadaraya-Watson estimator
\[
\hat m_N(w)
:=
\frac{\hat\Psi_N(w)}{\hat f_N(w)}
=
\frac{\sum_{\mathbf i_p\in\mathcal P_{p,N}}K_{\mathbf i_p}(w)Y_{\mathbf i_p}}
{\sum_{\mathbf i_p\in\mathcal P_{p,N}}K_{\mathbf i_p}(w)}.
\]

\begin{example}[Conditional subgraph probability]\label{ex:subgraph}
The leading example concerns the conditionally independent dyad (CID)
network formation model, in which agents \(i\ne j\) are linked
according to the binary variable
\(D_{ij}:=\mathbf1\{V_{\{i,j\}}\le g(X_i,X_j,U_i,U_j)\}\), and the links
are independent Bernoulli variables conditional on the primitives.
The graphon \(g\) is symmetric in its two agents, measurable, and takes
values in \([0,1]\).

Let \(G_N\) be the graph of $N$ agents
with adjacency matrix \((D_{ij})\). Define \(G_N[\mathbf i_p]\) to be the
subgraph induced by the $p$ agents in \(\mathbf i_p\), and let \(\simeq\)
denote graph isomorphism.

The polyadic
outcome of the running example
is the subgraph indicator
\(Y_{\mathbf i_p,S}:=\mathbf1\{G_N[\mathbf i_p]\simeq S\}\) for a fixed
graph \(S\) on \(p\) vertices, such as a triangle when $p=3$. Since each \(D_{i_ai_b}\) is a function of \(\Xi_{i_a}\),
\(\Xi_{i_b}\), and \(V_{\{i_a,i_b\}}\), the indicator has the form
\eqref{eq:polyadic-outcome}. Its regression function $m(w)$ is the conditional subgraph
probability
\[
P(S,w):= E[\mathbf1\{G_N[\mathbf i_p]\simeq S\}\mid X_{\mathbf i_p}=w] = P\bigl(G_N[\mathbf i_p]\simeq S\mid X_{\mathbf i_p}=w\bigr).
\]
Explicitly, conditional on \(X_{\mathbf i_p}=w\) and on the latent
variables \(U_1,\dots,U_p\) of the agents in positions \(1,\dots,p\),
agents \(a\) and \(b\) are linked independently with probability
\(g(w_a,w_b,U_a,U_b)\). Summing over the labeled graphs \(G\) on
\(\{1,\dots,p\}\) that are isomorphic to \(S\), with adjacency
indicators \(G_{ab}\), and integrating over the independent uniform
\(U_1,\dots,U_p\) gives
\[
P(S,w)
=
E\Bigl[\sum_{G\simeq S}\prod_{1\le a<b\le p}
g(w_a,w_b,U_a,U_b)^{G_{ab}}
\bigl(1-g(w_a,w_b,U_a,U_b)\bigr)^{1-G_{ab}}\Bigr].
\]
For instance, for a triangle,
\[
P(\text{triangle},w)
=E\bigl[g(w_1,w_2,U_1,U_2)\,g(w_1,w_3,U_1,U_3)\,g(w_2,w_3,U_2,U_3)\bigr].
\]
The estimator
\(\hat m_N(w)\), denoted \(\hat P_N(S,w)\), is
\[
\hat P_N(S,w)
=
\frac{\sum_{\mathbf i_p\in\mathcal P_{p,N}}
K_{\mathbf i_p}(w)\,\mathbf1\{G_N[\mathbf i_p]\simeq S\}}
{\sum_{\mathbf i_p\in\mathcal P_{p,N}}K_{\mathbf i_p}(w)},
\]
the fraction of \(p\)-tuples whose induced subgraph is
isomorphic to \(S\), in which each tuple is weighted by
\(K_{\mathbf i_p}(w)\), the kernel weight of its stacked covariates at
\(w\). Tuples whose covariates lie far from \(w\) receive little or no
weight, so \(\hat P_N(S,w)\) is a local version of the sample
frequency of \(S\) with covariates \(w\).
\end{example}

\section{Rates of convergence}\label{sec:rates}
This section derives the rates of convergence of
\(\hat m_N(w)\) in three steps.
Section~\ref{sec:variance} bounds the variance of the kernel numerator
at a fixed point. Section~\ref{sec:uniform-numerator} strengthens this
to a uniform bound over a region that expands at a polynomial rate.
Section~\ref{sec:uniform-regression} combines the uniform bounds for
the numerator and the denominator to
obtain the uniform rate of \(\hat m_N(w)\).

Two conventions for tuples are used throughout. First, set operations
on tuples (\(\cup\), \(\cap\), \(\setminus\), \(\subseteq\), \(\in\))
apply to the sets of their entries. Second, a set used as a subscript
selects the corresponding elements of a vector: for a set \(A\) of
agents, \(X_A\) and \(x_A\) stack the covariates of the agents in
\(A\), so that \(X_{\mathbf i_p\cup\mathbf j_p}\) collects the
covariates of the agents that appear in \(\mathbf i_p\) or
\(\mathbf j_p\). I begin with two mild regularity conditions.
\begin{assumption}[Regularity conditions]\label{aspt:Model} \text{}
\begin{enumerate}[label=(\roman*)]
\item\label{asptpart:density}
The covariate density is bounded: $\sup_{x\in\mathbb R^{d_X}}f(x)\le B_1<\infty.$
\item\label{asptpart:crossmoment}
For all \(\mathbf i_p,\mathbf j_p\in\mathcal P_{p,N}\)
that share at least one agent,
\[
\sup_{x_{\mathbf i_p\cup\mathbf j_p}}
E\!\left[
|Y_{\mathbf i_p}Y_{\mathbf j_p}|
\,\middle|\,
X_{\mathbf i_p\cup\mathbf j_p}
=x_{\mathbf i_p\cup\mathbf j_p}
\right]
\prod_{d\in\mathbf i_p\cup\mathbf j_p}f(x_d)
\le B_2<\infty.
\]
\end{enumerate}
\end{assumption}

Part~\ref{asptpart:density} is standard in kernel estimation.
Part~\ref{asptpart:crossmoment} bounds the cross moments of outcomes
on overlapping tuples, weighted by the joint density of the covariates
involved.

\begin{example*}[Example~\ref{ex:subgraph} continued]
In Example~\ref{ex:subgraph}, part~\ref{asptpart:density} of
Assumption~\ref{aspt:Model} is a condition on the covariates and is
maintained. Because the outcome is a subgraph indicator with values in
\(\{0,1\}\), part~\ref{asptpart:crossmoment} holds with
\(B_2=\max\{1,B_1\}^{2p-1}\), since two tuples that share an agent
involve at most \(2p-1\) agents.
\end{example*}

\begin{assumption}[Kernel]\label{aspt:KernelA}
The kernel \(K:\mathbb R^{d_W}\to\mathbb R\) satisfies:
\begin{enumerate}[label=(\roman*)]
\item\label{asptpart:KernelA_1}
\(\int_{\mathbb R^{d_W}}|K(u)|\,du\le C_1<\infty\).

\item\label{asptpart:KernelA_2}
For every \(A\subseteq\{1,\dots,p\}\),
\(\sup_{u_A}\int|K(u)|\,du_{\{1,\dots,p\}\setminus A}\le C_2<\infty\).

\item\label{asptpart:KernelA_3}
\(\sup_{u\in\mathbb R^{d_W}}|K(u)|\le K_{\max}<\infty\).
\end{enumerate}
\end{assumption}

Parts~\ref{asptpart:KernelA_1} on integrability and~\ref{asptpart:KernelA_3} on boundedness are standard in kernel estimation. Part~\ref{asptpart:KernelA_2} requires that integrating \(|K|\) over
any subset of the \(p\) blocks of its argument leaves a bounded
function of the remaining blocks. In fact, it contains
parts~\ref{asptpart:KernelA_1} and~\ref{asptpart:KernelA_3} as the
cases \(A=\varnothing\) and \(A=\{1,\dots,p\}\). The proofs
use part~\ref{asptpart:KernelA_2} whenever they condition on the
covariates of some agents in a tuple and integrate out those of the
others.

\subsection{Variance bound}\label{sec:variance}

The first result bounds the variance of the kernel numerator at a
fixed point.

\begin{theorem}[Variance bound]\label{thm:Variance-bound}
Under Assumptions~\ref{aspt:Model} and~\ref{aspt:KernelA}, and the
bandwidth condition \(Nh_N^{d_X}\to\infty\), there exists a constant
\(M_0<\infty\) such that, for all sufficiently large \(N\) and all $w\in\mathbb R^{d_W}$,
\[
\var{\hat\Psi_N(w)}
\le\frac{M_0}{Nh_N^{d_X}}.
\]
\end{theorem}

The proof is in Appendix~\ref{app:variance}. The idea is to group the pairs of tuples in the variance by the
number \(k\) of agents they share. There are \(O(N^{2p-k})\) such
pairs, and the covariance of their kernel-weighted outcomes is of order
\(h_N^{-kd_X}\). After normalization, the pairs that share a single
agent dominate and give the bound.
Notice that the bound has an effective sample-size interpretation. About \(N^ph_N^{pd_X}\)
ordered tuples are within the bandwidth of \(w\),
so if their outcomes were independent, the variance of
\(\hat\Psi_N(w)\) would be of order \((N^ph_N^{pd_X})^{-1}\).
Theorem~\ref{thm:Variance-bound} gives \((Nh_N^{d_X})^{-1}\) instead:
the effective number of observations is the number of agents in the
kernel window, not the number of tuples.

\subsection{Uniform convergence of the kernel numerator}
\label{sec:uniform-numerator}

Uniform convergence of \(\hat\Psi_N(w)\) requires higher moments of the
outcomes and a smoothness condition on the kernel. Let \(\|\cdot\|\) denote the sup norm on \(\mathbb R^{d_W}\).

\begin{assumption}[Regularity conditions]\label{aspt:regularity}
Assume that:
\begin{enumerate}[label=(\roman*)]
\item\label{asptpart:regularity1}
For some \(s>2\),
\(E|Y_{\mathbf i_p}|^s\le B_3<\infty\),
and
$\sup_{x_{\mathbf i_p}}
E\left[
|Y_{\mathbf i_p}|^s
\mid X_{\mathbf i_p}=x_{\mathbf i_p}
\right]
\prod_{d\in\mathbf i_p}f(x_d)
\le B_{2,s}<\infty$
for all \(\mathbf i_p\in\mathcal P_{p,N}\).

\item\label{asptpart:regularity2}
For some \(\Lambda_1<\infty\) and \(L_K<\infty\), \(K(u)=0\) for
\(\|u\|>L_K\), and \(|K(u)-K(v)|\le\Lambda_1\|u-v\|\) for all
\(u,v\in\mathbb R^{d_W}\).
\end{enumerate}
\end{assumption}
Assumption~\ref{aspt:regularity}\ref{asptpart:regularity1} strengthens the moment conditions in Assumption~\ref{aspt:Model}.
Assumption~\ref{aspt:regularity}\ref{asptpart:regularity2}
coincides with the compactly supported case of Assumption 3 in
\citet{hansen_uniform_2008}. It
provides the envelope that the covering argument in the proof
requires: for all \(\|w_1-w_2\|\le\delta\le L_K\),
\(|K(w_2)-K(w_1)|\le\delta K^*(w_1)\), where
\(K^*(u):=\Lambda_1\mathbf1\{\|u\|\le2L_K\}\). Indeed, if
\(\|w_1\|>2L_K\), both \(K(w_1)\) and \(K(w_2)\) vanish, and otherwise
the Lipschitz bound applies. The envelope \(K^*\) is bounded,
integrable, and compactly supported, so it satisfies
Assumption~\ref{aspt:KernelA} as well.

Define
\(a_N:=(\ln N/(Nh_N^{d_X}))^{1/2}\).
The proof truncates the outcomes at a threshold \(\tau_N\), which must
be large enough for the truncation error to be negligible and small
enough for the truncated statistic to concentrate. Two sequences
describe the admissible range. Let \(\phi_N\to\infty\), for example
\(\phi_N=(\ln(\ln N))^2\ln N\), and define
\[
\underline\tau_N
:=
\max\left\{
a_N^{-1/(s-1)},
\;
\min\left[
(a_Nh_N^{pd_X})^{-1/(s-1)},
\;
(N^p\phi_N)^{1/s}
\right]
\right\},
\]
and
\[
\overline\tau_N
:=
\min\left\{
a_N^{-1},
\;
\min_{2\le k\le p}
\left(\frac N{\ln N}\right)^{k/2}
h_N^{(2k-1)d_X/2}
\right\}.
\]

For positive sequences \(x_N\) and \(y_N\), \(x_N\ll y_N\), or
equivalently \(y_N\gg x_N\), means that \(x_N/y_N\to0\), and
\(x_N\asymp y_N\) means that \(x_N/y_N\) is bounded away from zero and
infinity.

\begin{theorem}[Uniform convergence in probability]
\label{thm:Weak-Uniform-Convergence}
Under Assumptions~\ref{aspt:Model}, \ref{aspt:KernelA}, and
\ref{aspt:regularity}, and the bandwidth conditions
\((N/\ln N)h_N^{d_X}\to\infty\) and \(\underline\tau_N\ll\overline\tau_N\),
for any \(q>0\) and any sequence \(c_N\) with \(0<c_N\le N^q\),
\[
\sup_{\|w\|\le c_N}
\left|\hat\Psi_N(w)-E\hat\Psi_N(w)\right|
=O_p(a_N).
\]
\end{theorem}

The theorem establishes convergence in probability uniformly over any
ball whose radius grows at most polynomially, including a fixed ball.
The proof is in
Appendix~\ref{app:uniform}, in which I split \(\hat\Psi_N\) into a statistic in
which the outcomes are truncated at \(\tau_N\) and a residual. The
centered truncated statistic is the sum of a \(U\)-statistic of order
\(p\) in the agent primitives and a link-shock remainder, which has
mean zero given the agent primitives. Bernstein's inequality controls
the H\'ajek projection of the \(U\)-statistic, the concentration
inequalities of \citet{arcones_limit_1993} control its higher-order
canonical projections, and McDiarmid's bounded-differences inequality,
applied conditionally on the agent primitives, controls the
link-shock remainder. These bounds require \(\tau_N\) to be small. The
residual is controlled through its expected supremum or the probability
that it is nonzero, which requires \(\tau_N\) to be large. The condition
\(\underline\tau_N\ll\overline\tau_N\) ensures that a threshold
satisfying both requirements exists, and a covering argument as in
\citet{hansen_uniform_2008} extends the bounds from a fixed point to the
supremum over the region.

\begin{example*}[Example~\ref{ex:subgraph} continued]
In Example~\ref{ex:subgraph}, the outcome is a subgraph indicator bounded by $1$, which simplifies the conditions of Theorem~\ref{thm:Weak-Uniform-Convergence}.
Assumption~\ref{aspt:regularity}\ref{asptpart:regularity1} holds for
every \(s\), with \(B_3=1\) and \(B_{2,s}=\max\{1,B_1\}^p\). With
\(\tau_N>1\), the truncation in the proof has no effect, so the lower
restriction on \(\tau_N\) is not needed, and the upper restrictions in
\(\overline\tau_N\) reduce to
\[
\left(\frac N{\ln N}\right)^{k/2}
h_N^{(2k-1)d_X/2}\to\infty,
\]
for $k=2,\dots,p$. Since \(h_N\to0\), the strongest of these restrictions is attained
at \(k=p\), which is $(N/\ln N)h_N^{(2-1/p)d_X}\to\infty.$
For \(h_N\asymp(\ln N/N)^{1/(2\beta+d_X)}\), the bandwidth choice in
Theorem~\ref{thm:Weak-Uniform-Convergence-for-kernel-estimator}, this
condition holds if and only if \(\beta>(p-1)d_X/(2p)\). For triangles
with scalar covariates, the condition is
\((N/\ln N)h_N^{5/3}\to\infty\).
\end{example*}

\subsection{Uniform convergence of the Nadaraya-Watson estimator}
\label{sec:uniform-regression}

I now turn to the regression estimator
\(\hat m_N(w)=\hat\Psi_N(w)/\hat f_N(w)\). Theorem~\ref{thm:Weak-Uniform-Convergence}
controls the stochastic error of the numerator and the denominator. A smoothing bias arises because the kernel averages over a
neighborhood of \(w\); controlling it requires a higher-order kernel.

\begin{assumption}[Higher-order kernel]\label{aspt:kernelB}
The kernel satisfies
\(\int_{\mathbb R^{d_W}}K(u)\,du=1\),
and, for every multi-index
\(\eta=(\eta_1,\dots,\eta_{d_W})\in\mathbb Z_+^{d_W}\) with
\(1\le|\eta|<\beta\),
\(\int_{\mathbb R^{d_W}}u^\eta K(u)\,du=0\), where
\(u^\eta:=u_1^{\eta_1}\cdots u_{d_W}^{\eta_{d_W}}\).
Also,
\(\int_{\mathbb R^{d_W}}\|u\|^\beta|K(u)|\,du<\infty\).
\end{assumption}

I now give a uniform convergence result for the regression estimator over a sequence of expanding sets. With \(c_N\) as in Theorem~\ref{thm:Weak-Uniform-Convergence}, define $\delta_N:=\inf_{\|w\|\le c_N}f_W(w)$ and
$$a_N^*:=\left(\frac{\ln N}{Nh_N^{d_X}}\right)^{1/2}+h_N^\beta.$$

\begin{theorem}[Uniform convergence in probability of the Nadaraya-Watson estimator]
\label{thm:Weak-Uniform-Convergence-for-kernel-estimator}
Suppose that $f_W, m\in \Sigma(\beta, L)$, $m$ is bounded, \(\delta_N>0\), and \(\delta_N^{-1}a_N^*\to0\).
Under the assumptions of
Theorem~\ref{thm:Weak-Uniform-Convergence} and
Assumption~\ref{aspt:kernelB},
\[
\sup_{\|w\|\le c_N}|\hat m_N(w)-m(w)|
=
O_p(\delta_N^{-1}a_N^*).
\]
If, in addition, \(h_N\asymp(\ln N/N)^{1/(2\beta+d_X)}\) satisfies the
bandwidth conditions of Theorem~\ref{thm:Weak-Uniform-Convergence},
then
\[
\sup_{\|w\|\le c_N}|\hat m_N(w)-m(w)|
=
O_p\left(
\delta_N^{-1}
\left(\frac{\ln N}{N}\right)^{\beta/(2\beta+d_X)}
\right).
\]
\end{theorem}

The proof is in Appendix~\ref{app:regression}. Notice that the additional term \(\delta_N^{-1}\) comes from the estimated
density. Because \(f_W\) integrates to one,
\(\delta_N(2c_N)^{d_W}\le1\), so \(\delta_N\to0\) whenever
\(c_N\to\infty\), i.e., the density must decay somewhere on a growing
region. Theorem~\ref{thm:Weak-Uniform-Convergence-for-kernel-estimator} allows this because \(\delta_N\) enters the rate. For light-tailed
covariates, the decay is fast. For example, with
standard Gaussian covariates, \(\delta_N\) is of order
\(\exp(-d_Wc_N^2/2)\), and \(\delta_N^{-1}a_N^*\to0\) allows \(c_N\) to
grow at most like \((\ln N)^{1/2}\). The leading case is a
fixed \(c_N\) such that \(f_W\) is bounded away from zero on
\(\{w:\|w\|\le c_N\}\). Then \(\delta_N\) is a positive constant, and
the rate is \(O_p(a_N^*)\), which equals
\((\ln N/N)^{\beta/(2\beta+d_X)}\) for the balanced bandwidth.

\begin{example*}[Example~\ref{ex:subgraph} continued]
Theorem~\ref{thm:Weak-Uniform-Convergence-for-kernel-estimator} applies to the conditional subgraph probability in Example~\ref{ex:subgraph} when \(f\) and \(g(\cdot,\cdot,u,u')\) are smooth enough, uniformly in \((u,u')\), that \(f_W\) and \(P(S,\cdot)\) belong to \(\Sigma(\beta,L)\).
For
triangles with scalar covariates and \(\beta=2\), on a fixed ball on which \(f_W\) is bounded away from zero,
the balanced bandwidth gives $\hat P_N(S,w)$ the rate \((\ln N/N)^{2/5}\), the rate of a one-dimensional problem, although
\(P(S,\cdot)\) is a function of three covariates. With \(N\) i.i.d.\
observations of three regressors, the optimal rate would be
\((\ln N/N)^{2/7}\).
\end{example*}

\section{Pointwise asymptotic normality}\label{sec:pointwise}

In this section, I prove a pointwise asymptotic normality result for the estimator at the rate
\((Nh_N^{d_X})^{1/2}\). Since the estimator \(\hat m_N(w)=\hat\Psi_N(w)/\hat f_N(w)\) is a ratio of
two random quantities built from the same kernel weights
on the same tuples, it is convenient to treat the numerator and denominator jointly. Let
\[
        \hat A_N(w):=
        \begin{pmatrix}\hat\Psi_N(w)\\[2pt]\hat f_N(w)\end{pmatrix},
        \qquad
        A(w):=
        \begin{pmatrix}\Psi(w)\\[2pt]f_W(w)\end{pmatrix},
\]
so that \(\hat m_N(w)=r(\hat A_N(w))\) and \(m(w)=r(A(w))\) with $r(a,b):= a/b$.
The
proof uses the H\'ajek reduction of \(\hat A_N(w)\) as a sum of the first
Hoeffding projections over agents. For agent primitives \(\xi\) and any
\(\mathbf i_p\in\mathcal P_{p,N}\), define
\[
h_{N,1}(\xi;w)
:=
\frac1p\sum_{a=1}^p
E\!\left[
K_{\mathbf i_p}(w)
\begin{pmatrix}Y_{\mathbf i_p}\\ 1\end{pmatrix}
\,\middle|\,
\Xi_{i_a}=\xi
\right]
-
E\!\left[
K_{\mathbf i_p}(w)
\begin{pmatrix}Y_{\mathbf i_p}\\ 1\end{pmatrix}
\right].
\]
Lemma~\ref{lem:pointwise-hajek} in Appendix~\ref{app:pointwise} shows
\begin{equation}\label{eq:hajek-heuristic}
\hat A_N(w)-E\hat A_N(w)
=
\frac pN\sum_{i=1}^Nh_{N,1}(\Xi_i;w)
+o_p\bigl((Nh_N^{d_X})^{-1/2}\bigr).
\end{equation}
Conditioning on one agent localizes only the \(d_X\) coordinates of its
covariates, so \(\var{h_{N,1}(\Xi_1;w)}\) is of order \(h_N^{-d_X}\)
(coinciding with the \(k=1\) bound in Theorem~\ref{thm:Variance-bound}). Hence, the leading
term in \eqref{eq:hajek-heuristic} has covariance of order
\((Nh_N^{d_X})^{-1}\). The next assumption asks that this covariance has a limit.

\begin{assumption}[Limiting first-projection variance]
\label{aspt:pointwise-projection}
At the fixed point \(w\) with \(f_W(w)>0\), the limit
\(\Omega(w):=\lim_{N\to\infty}p^2h_N^{d_X}\var{h_{N,1}(\Xi_1;w)}<\infty\). And
\(\sigma_m^2(w):=\nabla r(A(w))'\Omega(w)\nabla r(A(w))>0\), where
\(\nabla r(A(w))=(f_W(w)^{-1},\,-m(w)f_W(w)^{-1})'\).
\end{assumption}

\begin{theorem}[Pointwise asymptotic distribution]
\label{thm:kernel-ratio-clt}
Fix \(w\) with \(f_W(w)>0\), and suppose \(f_W\) and \(m\) belong to
\(\Sigma(\beta,L)\) on an open neighborhood of \(w\).
Suppose Assumptions~\ref{aspt:Model}, \ref{aspt:KernelA},
\ref{aspt:regularity}\ref{asptpart:regularity1}, \ref{aspt:kernelB},
and \ref{aspt:pointwise-projection} hold, and
\(Nh_N^{d_X}\to\infty\).
Let
\(m_N^0(w):=E\hat\Psi_N(w)/E\hat f_N(w)\).
Then
\[
\sqrt{Nh_N^{d_X}}
\bigl(\hat m_N(w)-m_N^0(w)\bigr)
\to_d N(0,\sigma_m^2(w)).
\]
If, in addition,
\(\sqrt{Nh_N^{d_X}}\,h_N^\beta\to0\),
then
\[
\sqrt{Nh_N^{d_X}}
\bigl(\hat m_N(w)-m(w)\bigr)
\to_d N(0,\sigma_m^2(w)).
\]
\end{theorem}

The proof, in Appendix~\ref{app:pointwise}, has four steps.
\begin{enumerate}[label=(\roman*)]
\item \emph{H\'ajek reduction.} Lemma~\ref{lem:pointwise-hajek} proves
\eqref{eq:hajek-heuristic}.
\item \emph{Central limit theorem for $\hat A_N(w)$.} The leading term in \eqref{eq:hajek-heuristic} is a sum of
\(N\) i.i.d.\ vectors with covariance converging to \(\Omega(w)\) by
Assumption~\ref{aspt:pointwise-projection}. The Lyapunov condition
\[
\sum_{i=1}^N
E\bigl\|p\sqrt{h_N^{d_X}/N}\,h_{N,1}(\Xi_i;w)\bigr\|^s
=O\bigl((Nh_N^{d_X})^{1-s/2}\bigr)\to0
\]
follows from the \(s\)-th moment bound in
Assumption~\ref{aspt:regularity}\ref{asptpart:regularity1} and the
partial-integration bound of
Assumption~\ref{aspt:KernelA}\ref{asptpart:KernelA_2}, so the
Lindeberg--Feller theorem gives
\(\sqrt{Nh_N^{d_X}}(\hat A_N(w)-E\hat A_N(w))\to_dN(0,\Omega(w))\).
\item \emph{Delta method.} Since \(f_W(w)>0\), \(\hat f_N(w)\) is
positive with probability tending to one, and the delta method applied
to \(r\) gives the first conclusion: asymptotic normality of
$\hat m_N(w)$ centered at $m_N^0(w)$, with variance
\(\nabla r(A(w))'\Omega(w)\nabla r(A(w))=\sigma_m^2(w)\).
\item \emph{Undersmoothing.} Lemma~\ref{lem:bias} gives the bias calculation
\(E\hat A_N(w)-A(w)=O(h_N^\beta)\) and hence
\(m_N^0(w)-m(w)=O(h_N^\beta)\). Under
\(\sqrt{Nh_N^{d_X}}\,h_N^\beta\to0\) this is negligible at the
\((Nh_N^{d_X})^{-1/2}\) scale, which gives the second conclusion of asymptotic normality of $\hat m_N(w)$ centered at $m(w)$.
\end{enumerate}

\section{Conclusion}\label{sec:conclusion}

This paper studies nonparametric regression for polyadic outcomes,
which are defined on groups of a fixed number of agents. I establish a
variance bound, uniform convergence rates in probability, and
pointwise asymptotic normality for the subgraph analogue of the
Nadaraya-Watson kernel regression estimator. The framework nests
dyadic regression. Its leading application is the conditional probability that a group of agents forms a given
subgraph, such as a triangle, but the results apply to any polyadic
outcome.
Two questions remain open. The minimax lower bound of
\citet{graham_minimax_2021} has not been extended to \(p\ge3\). Also, Theorem~\ref{thm:kernel-ratio-clt} is only a starting point for inference, and consistent variance estimation needs further study.

\clearpage
\appendix

\section{Notation}
\label{app:notation}

Extending the notation of Section~\ref{sec:framework} to tuples of
length \(r\le p\), let \(\mathcal P_{r,N}\) be the set of ordered
\(r\)-tuples of distinct elements of \([N]\), and let
\(\mathcal C_{r,N}:=\{(q_1,\ldots,q_r):1\le q_1<\cdots<q_r\le N\}\) be the set of unordered \(r\)-tuples.
For \(\mathbf q_p\in\mathcal C_{p,N}\), let
\(\mathcal P(\mathbf q_p)\subset\mathcal P_{p,N}\) be the set of its
\(p!\) orderings.
Throughout the appendix, \(C\) denotes a generic positive constant
that depends on neither \(N\) nor \(w\) and whose value may change from
line to line.

\section{Proof of Theorem~\ref{thm:Variance-bound}}
\label{app:variance}

\begin{proof}
Grouping the ordered tuples by the set of their entries gives
\[
\hat\Psi_N(w)
=
\binom Np^{-1}
\sum_{\mathbf q_p\in\mathcal C_{p,N}}
Z_{N,\mathbf q_p}(w),
\]
with $Z_{N,\mathbf q_p}(w)
:=
\frac1{p!}
\sum_{\mathbf i_p\in\mathcal P(\mathbf q_p)}
K_{\mathbf i_p}(w)Y_{\mathbf i_p}.$

Hence
\[
\var{\hat\Psi_N(w)}
=
\binom Np^{-2}
\sum_{\mathbf q_p\in\mathcal C_{p,N}}
\sum_{\mathbf r_p\in\mathcal C_{p,N}}
\Cov\bigl(Z_{N,\mathbf q_p}(w),Z_{N,\mathbf r_p}(w)\bigr).
\]

For \(k=0,1,\dots,p\), define the overlap class
\[
\mathcal I_{N,k}
:=
\left\{
(\mathbf q_p,\mathbf r_p)\in\mathcal C_{p,N}^2:
|\mathbf q_p\cap\mathbf r_p|=k
\right\}.
\]
Then
\[
\var{\hat\Psi_N(w)}
=
\binom Np^{-2}
\sum_{k=0}^p
\sum_{(\mathbf q_p,\mathbf r_p)\in\mathcal I_{N,k}}
\Cov\bigl(Z_{N,\mathbf q_p}(w),Z_{N,\mathbf r_p}(w)\bigr).
\]

If \(k=0\), the two tuples depend on disjoint collections of agent
primitives and link shocks. Consequently,
\(\Cov\bigl(Z_{N,\mathbf q_p}(w),Z_{N,\mathbf r_p}(w)\bigr)=0\).
Only the classes \(k=1,\dots,p\) contribute to the variance.

For \(1\le k\le p-1\), fix
\((\mathbf q_p,\mathbf r_p)\in\mathcal I_{N,k}\).
By \eqref{eq:exchangeable}, I may relabel the agents and assume
\(\mathbf q_p=(q_1,\dots,q_p)\) and \(\mathbf
r_p=(q_1,\dots,q_k,r_{k+1},\dots,r_p)\).
Then
\[
\left|\Cov\bigl(Z_{N,\mathbf q_p}(w),Z_{N,\mathbf r_p}(w)\bigr)\right|
\le
E\left|Z_{N,\mathbf q_p}(w)Z_{N,\mathbf r_p}(w)\right|
+
\left|EZ_{N,\mathbf q_p}(w)\right|\left|EZ_{N,\mathbf r_p}(w)\right|.
\]
I bound these two terms separately.

First, I show that
\(\sup_{w\in\mathbb R^{d_W}} \left|EZ_{N,\mathbf q_p}(w)\right|\) is bounded.
By the triangle inequality,
\[
\left|EZ_{N,\mathbf q_p}(w)\right|
\le
\frac1{p!}
\sum_{\mathbf i_p\in\mathcal P(\mathbf q_p)}
E\left|K_{\mathbf i_p}(w)Y_{\mathbf i_p}\right|.
\]
Fix \(\mathbf i_p\in\mathcal P(\mathbf q_p)\). Conditioning on
\(X_{\mathbf i_p}\) and writing \(x_{\mathbf i_p}\) for a value of
\(X_{\mathbf i_p}\) gives
\[
E\left|K_{\mathbf i_p}(w)Y_{\mathbf i_p}\right|
=
h_N^{-d_W}
\int
\left|K\!\left(\frac{x_{\mathbf i_p}-w}{h_N}\right)\right|
m_1(x_{\mathbf i_p})
\prod_{d\in\mathbf q_p}f(x_d)\,dx_{\mathbf q_p},
\]
where
\(m_r(x_{\mathbf i_p}):=E[|Y_{\mathbf i_p}|^r\mid X_{\mathbf i_p}=x_{\mathbf i_p}]\)
denotes the conditional absolute moment of order \(r\).
By Jensen's inequality, \(m_1\le m_2^{1/2}\), so
\[
m_1(x_{\mathbf i_p})\prod_{d\in\mathbf q_p}f(x_d)
\le
\Bigl(m_2(x_{\mathbf i_p})\prod_{d\in\mathbf q_p}f(x_d)\Bigr)^{1/2}
\Bigl(\prod_{d\in\mathbf q_p}f(x_d)\Bigr)^{1/2}
\le
B_2^{1/2}B_1^{p/2},
\]
where the last step applies
Assumption~\ref{aspt:Model} with both tuples equal
to \(\mathbf i_p\) and the density bound \(f\le B_1\).
It follows that
\[
E\left|K_{\mathbf i_p}(w)Y_{\mathbf i_p}\right|
\le
B_2^{1/2}B_1^{p/2}h_N^{-d_W}
\int
\left|K\!\left(\frac{x_{\mathbf i_p}-w}{h_N}\right)\right|
\,dx_{\mathbf q_p}.
\]
The change of variables
\(u=(x_{\mathbf i_p}-w)/h_N\), with \(dx_{\mathbf q_p}=h_N^{d_W}\,du\),
and Assumption~\ref{aspt:KernelA}\ref{asptpart:KernelA_1} give
\[
h_N^{-d_W}
\int
\left|K\!\left(\frac{x_{\mathbf i_p}-w}{h_N}\right)\right|
\,dx_{\mathbf q_p}
=
\int_{\mathbb R^{d_W}}|K(u)|\,du
\le C_1.
\]
Thus
\(E\left|K_{\mathbf i_p}(w)Y_{\mathbf i_p}\right| \le
B_2^{1/2}B_1^{p/2}C_1\).
This bound is uniform in \(w\), \(N\), and \(\mathbf i_p\), so
\(\sup_{w\in\mathbb R^{d_W}}|EZ_{N,\mathbf q_p}(w)|\) is bounded, as
desired.
As a result, since \(h_N\to0\), for all sufficiently large \(N\),
\[
\left|EZ_{N,\mathbf q_p}(w)\right|\left|EZ_{N,\mathbf r_p}(w)\right|
\le Ch_N^{-kd_X}.
\]

Second,
\[
E\left|Z_{N,\mathbf q_p}(w)Z_{N,\mathbf r_p}(w)\right|
\le
\frac1{(p!)^2}
\sum_{\mathbf i_p\in\mathcal P(\mathbf q_p)}
\sum_{\mathbf j_p\in\mathcal P(\mathbf r_p)}
E\left|K_{\mathbf i_p}(w)Y_{\mathbf i_p}K_{\mathbf j_p}(w)Y_{\mathbf j_p}\right|.
\]
For fixed \(\mathbf i_p,\mathbf j_p\),
\begin{align*}
&E\left|
K_{\mathbf i_p}(w)Y_{\mathbf i_p}
K_{\mathbf j_p}(w)Y_{\mathbf j_p}
\right|\\
&\quad=
h_N^{-2pd_X}
\int
E\!\left[
|Y_{\mathbf i_p}Y_{\mathbf j_p}|
\,\middle|\,
X_{\mathbf q_p\cup\mathbf r_p}=x_{\mathbf q_p\cup\mathbf r_p}
\right]
\left|K\!\left(\frac{x_{\mathbf i_p}-w}{h_N}\right)\right|
\left|K\!\left(\frac{x_{\mathbf j_p}-w}{h_N}\right)\right|
\prod_{d\in\mathbf q_p\cup\mathbf r_p}f(x_d)\,dx_d\\
&\quad\le
B_2h_N^{-2pd_X}
\int
\left|K\!\left(\frac{x_{\mathbf i_p}-w}{h_N}\right)\right|
\left|K\!\left(\frac{x_{\mathbf j_p}-w}{h_N}\right)\right|
\,dx_{\mathbf q_p\cup\mathbf r_p}\\
&\quad\le
B_2h_N^{-kd_X}
\int
|K(u_{\mathbf q_p})|\,|K(u_{\mathbf r_p})|
\,du_{\mathbf q_p}\,du_{\mathbf r_p\setminus\mathbf q_p},
\end{align*}
where the first inequality is by Assumption~\ref{aspt:Model}\ref{asptpart:crossmoment} and the second by the change of variables in the covariates of the \(2p-k\) distinct agents, which yields
a factor \(h_N^{(2p-k)d_X}\).
Lastly, the remaining integral $\int
|K(u_{\mathbf q_p})|\,|K(u_{\mathbf r_p})|
\,du_{\mathbf q_p}\,du_{\mathbf r_p\setminus\mathbf q_p}$ is bounded by \(C_2C_1\), uniformly in
\(w,\mathbf i_p,\mathbf j_p\), which holds by integrating the \(p-k\) coordinates corresponding to
  \(\mathbf r_p\setminus\mathbf q_p\) in the second kernel and then applying
  Assumptions~\ref{aspt:KernelA}\ref{asptpart:KernelA_2} and
  \ref{aspt:KernelA}\ref{asptpart:KernelA_1}. Hence,
$$E\left|Z_{N,\mathbf q_p}(w)Z_{N,\mathbf r_p}(w)\right| \le Ch_N^{-kd_X}.$$

Combining the two bounds proves that for $1\le k\le p-1$,
\begin{equation}\label{eq:Vk-bound}
\sup_{(\mathbf q_p,\mathbf r_p)\in\mathcal I_{N,k}}
\left|
\Cov\bigl(Z_{N,\mathbf q_p}(w),Z_{N,\mathbf r_p}(w)\bigr)
\right|
\le Ch_N^{-kd_X}.
\end{equation}

For the diagonal class \(k=p\),
$\sup_{\mathbf q_p\in\mathcal C_{p,N}}
\var{Z_{N,\mathbf q_p}(w)}
\le
\sup_{\mathbf q_p\in\mathcal C_{p,N}}
E\bigl[Z_{N,\mathbf q_p}(w)^2\bigr].$
Using
\((\sum_{m=1}^{p!}a_m)^2\le p!\sum_{m=1}^{p!}a_m^2\), exchangeability
as in \eqref{eq:exchangeable}, and
Assumption~\ref{aspt:KernelA}\ref{asptpart:KernelA_3},
\begin{align*}
E\bigl[Z_{N,\mathbf q_p}(w)^2\bigr]
&\le
\frac1{p!}
\sum_{\mathbf i_p\in\mathcal P(\mathbf q_p)}
E\left[K_{\mathbf i_p}(w)^2Y_{\mathbf i_p}^2\right]\\
&\le
h_N^{-2pd_X}
\int
E\!\left[
Y_{\mathbf q_p}^2
\,\middle|\,
X_{\mathbf q_p}=x_{\mathbf q_p}
\right]
K\!\left(\frac{x_{\mathbf q_p}-w}{h_N}\right)^2
\prod_{d\in\mathbf q_p}f(x_d)\,dx_d\\
&\le Ch_N^{-pd_X},
\end{align*}
where the last inequality uses \(K^2\le K_{\max}|K|\),
Assumption~\ref{aspt:Model}\ref{asptpart:crossmoment} with
\(\mathbf i_p=\mathbf j_p\), and the change of variables above, so that
\(C=B_2C_1K_{\max}\). Therefore,
\begin{equation}\label{eq:Vp-bound}
\sup_{\mathbf q_p\in\mathcal C_{p,N}}
\var{Z_{N,\mathbf q_p}(w)}
\le Ch_N^{-pd_X}.
\end{equation}

For any fixed \(\mathbf q_p\in\mathcal C_{p,N}\), the number of
\(\mathbf r_p\in\mathcal C_{p,N}\) sharing exactly \(k\) agents is
\(\binom pk\binom{N-p}{p-k}\).
Thus
\(|\mathcal I_{N,k}| = \binom Np\binom pk\binom{N-p}{p-k}\).
Using \eqref{eq:Vk-bound} and \eqref{eq:Vp-bound}, $\var{\hat\Psi_N(w)}$ is bounded above by
\begin{equation*}
\binom Np^{-2}
\sum_{k=1}^{p-1}
|\mathcal I_{N,k}|
\sup_{(\mathbf q_p,\mathbf r_p)\in\mathcal I_{N,k}}
\left|
\Cov\bigl(Z_{N,\mathbf q_p}(w),Z_{N,\mathbf r_p}(w)\bigr)
\right|+
\binom Np^{-2}|\mathcal I_{N,p}|
\sup_{\mathbf q_p\in\mathcal C_{p,N}}
\var{Z_{N,\mathbf q_p}(w)},
\end{equation*}
which is bounded by $C\binom Np^{-1}
\sum_{k=1}^p
\binom pk\binom{N-p}{p-k}h_N^{-kd_X}$.
Since \(p\) is fixed, for $k=1,\dots,p$
\[
\frac{\binom{N-p}{p-k}}{\binom Np}
=O(N^{-k}).
\]
It follows that
\[
\var{\hat\Psi_N(w)}
\le
C\sum_{k=1}^pN^{-k}h_N^{-kd_X}
=
\frac C{Nh_N^{d_X}}
\sum_{j=0}^{p-1}(Nh_N^{d_X})^{-j}.
\]
The bandwidth condition \(Nh_N^{d_X}\to\infty\) bounds the last sum
uniformly for all sufficiently large \(N\). Hence there exists
\(M_0<\infty\) such that
\(\var{\hat\Psi_N(w)}\le M_0/(Nh_N^{d_X})\) for all \(w\in\mathbb
R^{d_W}\),
as claimed.
\end{proof}

\section{Proof of Theorem~\ref{thm:Weak-Uniform-Convergence}}
\label{app:uniform}

\subsection{Truncation and decomposition}\label{sec:trunc_decomp}

Let \(\tau_N\) be a truncation threshold with \(\underline\tau_N\ll\tau_N\ll\overline\tau_N\), which exists by the bandwidth condition \(\underline\tau_N\ll\overline\tau_N\), and decompose
\(\hat\Psi_N(w)=\widetilde\Psi_N(w)+R_N(w)\), where
\[
\widetilde\Psi_N(w)
=
\frac1{M_N}
\sum_{\mathbf i_p\in\mathcal P_{p,N}}
K_{\mathbf i_p}(w)Y_{\mathbf i_p}
\mathbf1\{|Y_{\mathbf i_p}|<\tau_N\}
\]
is the truncated statistic and \(R_N(w)\) is the residual. Define
\[
\widetilde Z_{N,\mathbf q_p}(w)
:=
\frac1{p!}
\sum_{\mathbf i_p\in\mathcal P(\mathbf q_p)}
K_{\mathbf i_p}(w)Y_{\mathbf i_p}
\mathbf1\{|Y_{\mathbf i_p}|<\tau_N\},
\]
so that
\[
\widetilde\Psi_N(w)
=
\binom Np^{-1}
\sum_{\mathbf q_p\in\mathcal C_{p,N}}
\widetilde Z_{N,\mathbf q_p}(w).
\]

The statistic \(\widetilde\Psi_N\) is not a \(U\)-statistic, because \(\widetilde Z_{N,\mathbf q_p}(w)\) depends
not only on the primitives of the agents in \(\mathbf q_p\) but also on
the link shocks among them, which tuples sharing two or more agents
have in common. Let
\(\Xi_{\mathbf q_p}:=(\Xi_q:q\in\mathbf q_p)\). To separate the
two sources of randomness, average out the link shocks and define the projected kernel:
\[
\overline Z_{N,\mathbf q_p}(w)
:=
E\bigl[\widetilde Z_{N,\mathbf q_p}(w)\,\big|\,\Xi_{\mathbf q_p}\bigr].
\]
 Then, I use the following decomposition
\[
\widetilde\Psi_N(w)-E\widetilde\Psi_N(w)
=
U_N(w)+V_N(w),
\]
where
\[
U_N(w)
:=
\binom Np^{-1}
\sum_{\mathbf q_p\in\mathcal C_{p,N}}
\left(
\overline Z_{N,\mathbf q_p}(w)
-
E\overline Z_{N,\mathbf q_p}(w)
\right)
\]
and
\[
V_N(w)
:=
\binom Np^{-1}
\sum_{\mathbf q_p\in\mathcal C_{p,N}}
\left(
\widetilde Z_{N,\mathbf q_p}(w)
-
\overline Z_{N,\mathbf q_p}(w)
\right).
\]
The term \(U_N\) now is an order-\(p\) \(U\)-statistic in the i.i.d.\
agent primitives. The term \(V_N\) is the conditionally mean-zero
remainder from link shocks.

Let
\(\theta_N(w) := E[\overline Z_{N,\{1,\dots,p\}}(w)]\).
Define the canonical Hoeffding kernels recursively by
\[
\tilde h_{N,1}(\xi_1;w)
:=
E[\overline Z_{N,\{1,\dots,p\}}(w)\mid\Xi_1=\xi_1]
-\theta_N(w),
\]
and, for \(2\le k\le p\),
\[
\tilde h_{N,k}(\xi_1,\dots,\xi_k;w)
:=
E[\overline Z_{N,\{1,\dots,p\}}(w)
\mid\Xi_1=\xi_1,\dots,\Xi_k=\xi_k]
-\theta_N(w)-
\sum_{j=1}^{k-1}
\sum_{\substack{A\subseteq\{1,\dots,k\}\\|A|=j}}
\tilde h_{N,j}(\xi_A;w).
\]
Then
\[
U_N(w)=\sum_{k=1}^pU_{N,k}(w),
\qquad
U_{N,k}(w)
:=
\frac{\binom pk}{\binom Nk}
\sum_{\mathbf q_k\in\mathcal C_{k,N}}
\tilde h_{N,k}(\Xi_{\mathbf q_k};w).
\]
In particular, the first-order term (the H\'ajek projection) is
$$U_{N,1}(w)=\frac pN\sum_{j=1}^N\tilde h_{N,1}(\Xi_j;w).$$
It is an average of i.i.d.\ mean-zero random
variables, whereas $U_{N,k}(w)$ for \(k\ge2\) is a degenerate $k$th-order
$U$-statistic:
\[
E[\tilde h_{N,k}(\Xi_1,\dots,\Xi_k;w)
\mid\Xi_1,\dots,\Xi_{k-1}]
=0
\qquad\text{a.s.}
\]

\subsection{Concentration inequality and truncation bounds}\label{sec:concentration}

For the proof, I use the following lemma that supplies the concentration and truncation bounds for the terms in subsection~\ref{sec:trunc_decomp}.

\begin{lemma}\label{lem:uniformconv}
Suppose Assumptions~\ref{aspt:Model} and~\ref{aspt:KernelA} hold. For any \(\alpha>0\), there exists
\(M_\alpha<\infty\) such that:
\begin{enumerate}[label=(\roman*)]
\item\label{lemmapart:polytailbound1}
If \(\tau_N\ll a_N^{-1}\), then
$\sup_{w\in\mathbb R^{d_W}}
P\bigl(|U_{N,1}(w)|>M_\alpha a_N\bigr)
=O(N^{-\alpha}).$

\item\label{lemmapart:polytailbound2}
If
$\tau_N
\ll
\min_{2\le k\le p}
\left(\frac N{\ln N}\right)^{k/2}
h_N^{(2k-1)d_X/2}$
and \(a_N=o(1)\), then, for every \(2\le k\le p\),
\[
\sup_{w\in\mathbb R^{d_W}}
P\bigl(|U_{N,k}(w)|>M_\alpha a_N\bigr)
=O(N^{-\alpha}).
\]

\item\label{lemmapart:residual1}
If, for some \(s>1\), the density-weighted conditional moment bound of
Assumption~\ref{aspt:regularity}\ref{asptpart:regularity1} holds with
constant \(B_{2,s}\), and \(\tau_N\gg a_N^{-1/(s-1)}\), then
\[
\sup_{w\in\mathbb R^{d_W}}
\left|E[\hat\Psi_N(w)-\widetilde\Psi_N(w)]\right|
=o(a_N).
\]

\item\label{lemmapart:residual2}
If, for some \(s>1\),
\(E|Y_{\mathbf i_p}|^s\le B_3\) and
\(\tau_N\gg(a_Nh_N^{pd_X})^{-1/(s-1)}\),
then
\[
\sup_{w\in\mathbb R^{d_W}}
|\hat\Psi_N(w)-\widetilde\Psi_N(w)|
=o_p(a_N).
\]

\item\label{lemmapart:equalitywpa1}
If, for some \(s>1\),
\(E|Y_{\mathbf i_p}|^s\le B_3\) and
\(\tau_N\ge(N^p\phi_N)^{1/s}\) with \(\phi_N=(\ln(\ln N))^2\ln N\), then
\[
P(\hat\Psi_N=\widetilde\Psi_N)
=
P\bigl(
\hat\Psi_N(w)=\widetilde\Psi_N(w)
\text{ for all }w\in\mathbb R^{d_W}
\bigr)
\to1.\footnote{This is an equality of functions, but the event is
measurable: \(\hat\Psi_N-\widetilde\Psi_N\) equals
\(M_N^{-1}\sum_{\mathbf i_p}K_{\mathbf i_p}(\cdot)
Y_{\mathbf i_p}\mathbf1\{|Y_{\mathbf i_p}|\ge\tau_N\}\), a finite sum
of continuous functions of \(w\) with random coefficients, so it
vanishes identically if and only if it vanishes on a countable dense
set of \(w\), a countable intersection of measurable events.}
\]

\item\label{lemmapart:linkshock}
If \(\tau_N\ll a_N^{-1}\), \(a_N=o(1)\), and \(K(u)=0\) for
\(\|u\|>L_0\), for some \(L_0<\infty\), then
\[
\sup_{w\in\mathbb R^{d_W}}
P\bigl(|V_N(w)|>M_\alpha a_N\bigr)
=O(N^{-\alpha}).
\]
\end{enumerate}
\end{lemma}

Lemma~\ref{lem:uniformconv} is proved in
Appendix~\ref{appendixsection:lemmaforuniformconv}.

\subsection{Proof of the theorem}

\begin{proof}[Proof of Theorem~\ref{thm:Weak-Uniform-Convergence}]
The bandwidth condition
\(\underline\tau_N\ll\overline\tau_N\) ensures that a truncation threshold \(\tau_N\) satisfying the following conditions exists:
\begin{enumerate}[label=\arabic*.]
\item \(\tau_N\ll a_N^{-1}\);
\item
\(\tau_N\ll\min_{2\le k\le p}(N/\ln N)^{k/2}h_N^{(2k-1)d_X/2}\);
\item \(\tau_N\gg a_N^{-1/(s-1)}\);
\item
\(\tau_N\ge(N^p\phi_N)^{1/s}\) or
\(\tau_N\gg(a_Nh_N^{pd_X})^{-1/(s-1)}\).
\end{enumerate}

Let \(\alpha:=(q+1/2)d_W+p+1\). Let \(M_\alpha\) be the constant of
Lemma~\ref{lem:uniformconv} for this \(\alpha\), chosen so that the
lemma holds both for \(\widetilde\Psi_N\) and for the envelope statistic
\(\breve\Psi_N\) (defined below), and fix
\(M\ge\max\{M_\alpha,B_2^{1/2}B_1^{p/2}\int|K^*(w)|\,dw\}\).

For any \(\mathcal W_N\subset\mathbb R^{d_W}\),
\begin{align*}
&P\left(
\sup_{w\in\mathcal W_N}
|\hat\Psi_N(w)-E\hat\Psi_N(w)|
>(2p+6)Ma_N
\right)\\
&\quad\le
\underbrace{
P\left(
\sup_{w\in\mathcal W_N}
|\widetilde\Psi_N(w)-E\widetilde\Psi_N(w)|
>(2p+4)Ma_N
\right)
}_{(\star)}+
\underbrace{
P\left(
\sup_{w\in\mathcal W_N}
|R_N(w)-ER_N(w)|>2Ma_N
\right)
}_{(\star\star)}.
\end{align*}

First, the residual term $(\star\star)=o(1)$ because
\[
\begin{split}
(\star\star)
&\le
P\Bigl(\sup_{w\in\mathbb R^{d_W}}|R_N(w)-ER_N(w)|>2Ma_N\Bigr)\\
&\le
P\Bigl(\sup_{w\in\mathbb R^{d_W}}|R_N(w)|>Ma_N\Bigr)
+
\mathbf1\Bigl\{\sup_{w\in\mathbb R^{d_W}}|ER_N(w)|>Ma_N\Bigr\}.
\end{split}
\]
Part~\ref{lemmapart:residual1} of Lemma~\ref{lem:uniformconv}
  implies the indicator term becomes zero for sufficiently large \(N\). Also,
\[
P\left(
\sup_{w\in\mathbb R^{d_W}}|R_N(w)|>Ma_N
\right)\le
\min\left\{
1-P(\hat\Psi_N=\widetilde\Psi_N),
\frac{
E\sup_{w\in\mathbb R^{d_W}}|R_N(w)|
}{Ma_N}
\right\}.
\]
On the right-hand side, depending on which alternative in condition~4
holds, either Lemma~\ref{lem:uniformconv}
part~\ref{lemmapart:equalitywpa1} gives
\(P(\hat\Psi_N=\widetilde\Psi_N)\to1\), or
part~\ref{lemmapart:residual2} gives
\(E\sup_{w}|R_N(w)|=o(a_N)\).

Second, for \((\star)\), a covering argument reduces the supremum over
the region to a maximum over finitely many points, at which the
pointwise concentration bounds of Lemma~\ref{lem:uniformconv} apply.
This part closely follows \citet{hansen_uniform_2008} and
\citet{graham_minimax_2021}.
Cover a compact region \(\mathcal W_N\subset\mathbb R^{d_W}\) by
finitely many balls of radius \(a_Nh_N\). Denote their centers by
\(w_{N,1},\dots,w_{N,L_N}\), and for \(j=1,\dots,L_N\), let
\[
A_{N,j}:=\{w\in\mathbb R^{d_W}:\|w-w_{N,j}\|\le a_Nh_N\}.
\]
For sufficiently large \(N\), \(a_N<L_K\). For \(w\in A_{N,j}\),
Assumption~\ref{aspt:regularity}\ref{asptpart:regularity2} implies
\begin{equation}\label{eq:Lipschitz}
\left|
K\!\left(\frac{X_{\mathbf i_p}-w}{h_N}\right)
-
K\!\left(\frac{X_{\mathbf i_p}-w_{N,j}}{h_N}\right)
\right|
\le
a_NK^*\!\left(\frac{X_{\mathbf i_p}-w_{N,j}}{h_N}\right).
\end{equation}

Replacing \(K\) by \(K^*\) and \(Y_{\mathbf i_p}\) by \(|Y_{\mathbf i_p}|\) in \(\widetilde\Psi_N(w)\), I define the envelope statistic
\[
\breve\Psi_N(w)
:=
\frac1{M_N}
\sum_{\mathbf i_p\in\mathcal P_{p,N}}
|Y_{\mathbf i_p}|\mathbf1\{|Y_{\mathbf i_p}|<\tau_N\}
K^*_{\mathbf i_p}(w)
\]
with $K^*_{\mathbf i_p}(w)
:=
h_N^{-d_W}K^*\!\left(\frac{X_{\mathbf i_p}-w}{h_N}\right)$.
Since \(K^*\ge0\), \(\breve\Psi_N(w)\ge0\).

By the triangle inequality and \eqref{eq:Lipschitz}, for
\(w\in A_{N,j}\),
\[
\begin{split}
|\widetilde\Psi_N(w)-\widetilde\Psi_N(w_{N,j})|
&\le
\frac1{M_N}
\sum_{\mathbf i_p\in\mathcal P_{p,N}}
|Y_{\mathbf i_p}|\mathbf1\{|Y_{\mathbf i_p}|<\tau_N\}
h_N^{-d_W}
\left|
K\!\left(\frac{X_{\mathbf i_p}-w}{h_N}\right)
-
K\!\left(\frac{X_{\mathbf i_p}-w_{N,j}}{h_N}\right)
\right|\\
&\le a_N\breve\Psi_N(w_{N,j})
\end{split}
\]
with $0\le E\breve\Psi_N(w_{N,j})
\le
B_2^{1/2}B_1^{p/2}
\int|K^*(w)|\,dw<\infty$, by the first-moment bound in the proof of
Theorem~\ref{thm:Variance-bound} with \(K^*\) in place of \(K\).
Taking expectations gives
\(|E[\widetilde\Psi_N(w)-\widetilde\Psi_N(w_{N,j})]|
\le a_NE\breve\Psi_N(w_{N,j})\).
Thus, using these together with triangle inequalities,
\begin{align*}
&\sup_{w\in A_{N,j}}
|\widetilde\Psi_N(w)-E\widetilde\Psi_N(w)|\\
&\quad\le
|\widetilde\Psi_N(w_{N,j})-E\widetilde\Psi_N(w_{N,j})|+
\sup_{w\in A_{N,j}}
|\widetilde\Psi_N(w)-\widetilde\Psi_N(w_{N,j})|
+
\sup_{w\in A_{N,j}}
|E[\widetilde\Psi_N(w)-\widetilde\Psi_N(w_{N,j})]|\\
&\quad\le
|\widetilde\Psi_N(w_{N,j})-E\widetilde\Psi_N(w_{N,j})|
+
a_N\breve\Psi_N(w_{N,j})+a_N E\breve\Psi_N(w_{N,j})\\
&\quad\le
|\widetilde\Psi_N(w_{N,j})-E\widetilde\Psi_N(w_{N,j})|
+
a_N|\breve\Psi_N(w_{N,j})-E\breve\Psi_N(w_{N,j})|
+
2a_NE\breve\Psi_N(w_{N,j})\\
&\quad\le
|\widetilde\Psi_N(w_{N,j})-E\widetilde\Psi_N(w_{N,j})|
+
|\breve\Psi_N(w_{N,j})-E\breve\Psi_N(w_{N,j})|
+
2a_NE\breve\Psi_N(w_{N,j}),
\end{align*}
where the last inequality uses \(a_N\le1\).

By the choice of \(M\) mentioned at the beginning, \(E\breve\Psi_N(w_{N,j})\le M\), hence
\[
\begin{split}
&P\Bigl(\sup_{w\in A_{N,j}}
|\widetilde\Psi_N(w)-E\widetilde\Psi_N(w)|>(2p+4)Ma_N\Bigr)\\
&\quad\le
P\bigl(|\widetilde\Psi_N(w_{N,j})-E\widetilde\Psi_N(w_{N,j})|>(p+1)Ma_N\bigr)
+
P\bigl(|\breve\Psi_N(w_{N,j})-E\breve\Psi_N(w_{N,j})|>(p+1)Ma_N\bigr).
\end{split}
\]
Summing over the covering balls gives
\[
\begin{split}
(\star)
&\le
L_N\max_{1\le j\le L_N}
P\bigl(|\widetilde\Psi_N(w_{N,j})-E\widetilde\Psi_N(w_{N,j})|>(p+1)Ma_N\bigr)\\
&\quad+
L_N\max_{1\le j\le L_N}
P\bigl(|\breve\Psi_N(w_{N,j})-E\breve\Psi_N(w_{N,j})|>(p+1)Ma_N\bigr).
\end{split}
\]

The bounding argument is the same for these two quantities, as both $K$ and $K^*$ satisfy Assumption~\ref{aspt:KernelA}.
The truncation and decomposition in subsection~\ref{sec:trunc_decomp} and the bounds in subsection~\ref{sec:concentration} for $\widetilde\Psi_N(w_{N,j})$ also apply to \(\breve\Psi_N\), with outcome
\(|Y_{\mathbf i_p}|\) and kernel \(K^*\).
In particular, Lemma~\ref{lem:uniformconv} part~\ref{lemmapart:linkshock} applies to \(\breve\Psi_N\), because \(K^*\) vanishes outside \(\{\|u\|\le2L_K\}\).

Since \(\widetilde\Psi_N-E\widetilde\Psi_N=\sum_{k=1}^pU_{N,k}+V_N\) and
\(M\ge M_\alpha\), parts~\ref{lemmapart:polytailbound1},
\ref{lemmapart:polytailbound2}, and~\ref{lemmapart:linkshock} of
Lemma~\ref{lem:uniformconv} give
\[
\begin{split}
&\sup_{w\in\mathbb R^{d_W}}
P\bigl(|\widetilde\Psi_N(w)-E\widetilde\Psi_N(w)|>(p+1)Ma_N\bigr)\\
&\quad\le
\sum_{k=1}^p
\sup_{w\in\mathbb R^{d_W}}P\bigl(|U_{N,k}(w)|>Ma_N\bigr)
+
\sup_{w\in\mathbb R^{d_W}}P\bigl(|V_N(w)|>Ma_N\bigr)
=O(N^{-\alpha}),
\end{split}
\]
and the same bound holds for \(\breve\Psi_N\). These bounds are
uniform in \(w\), so they apply at every center \(w_{N,j}\),
\(j=1,\dots,L_N\). Hence
each of the two maxima over \(j\) in the bound on \((\star)\) is
\(O(N^{-\alpha})\), and \((\star)=O(L_NN^{-\alpha})\).

For \(\mathcal W_N=\{w:\|w\|\le c_N\}\) with \(c_N\le N^q\), the
number of covering balls can be taken as
\(L_N=C(c_N/(a_Nh_N))^{d_W}\). $L_N$ is bounded by a power
of \(N\). First, for large \(N\), \(Nh_N^{d_X}\ge\ln N\ge1\), so
\(a_N^{-1}=(Nh_N^{d_X}/\ln N)^{1/2}\le N^{1/2}\). Second,
\(h_N^{d_X}\ge\ln N/N\ge N^{-1}\) for large \(N\), so
\(h_N^{-d_X}\le N\) and hence \(h_N^{-d_W}=(h_N^{-d_X})^p\le N^p\).
Together with \(c_N\le N^q\),
\[
L_N
\le
C\,c_N^{d_W}\,a_N^{-d_W}\,h_N^{-d_W}
\le
C\,N^{qd_W}\,N^{d_W/2}\,N^p
=
C\,N^{(q+1/2)d_W+p}.
\]
The choice \(\alpha=(q+1/2)d_W+p+1\) exceeds this exponent, so
\(L_NN^{-\alpha}=O(N^{-1})=o(1)\).
Therefore, $(\star)=o(1)$.
\end{proof}

\section{Proof of Lemma \ref{lem:uniformconv}}
\label{appendixsection:lemmaforuniformconv}

\subsection{Part~\ref{lemmapart:polytailbound1}}

\begin{proof}[Proof of Lemma~\ref{lem:uniformconv},
part~\ref{lemmapart:polytailbound1}]
The proof applies Bernstein's inequality to the H\'ajek projection.
Recall that
\[
U_{N,1}(w)=\frac pN\sum_{j=1}^N\tilde h_{N,1}(\Xi_j;w),
\qquad
\tilde h_{N,1}(\xi_1;w)
=
E[\overline Z_{N,\{1,\dots,p\}}(w)\mid\Xi_1=\xi_1]
-\theta_N(w),
\]
with \(\theta_N(w)=E[\overline Z_{N,\{1,\dots,p\}}(w)]\). Define
\(Q_j(w):=\tau_N^{-1}h_N^{d_X}\tilde h_{N,1}(\Xi_j;w)\). For fixed
\(w\), the variables \(Q_1(w),\dots,Q_N(w)\) are i.i.d.\ with mean zero,
because the \(\Xi_j\) are i.i.d.\ and \(E[\tilde h_{N,1}(\Xi_1;w)]=0\).

\emph{Step 1: the summands are bounded.} Writing out
\(\overline Z_{N,\{1,\dots,p\}}(w)\) as the average over the \(p!\)
orderings \(\mathbf i_p\in\mathcal P(\{1,\dots,p\})\),
\begin{align*}
\tau_N^{-1}h_N^{d_X}
\bigl|E[\overline Z_{N,\{1,\dots,p\}}(w)\mid\Xi_j]\bigr|
&=
\tau_N^{-1}h_N^{d_X}
\Bigl|
E\Bigl[\frac1{p!}\sum_{\mathbf i_p}
Y_{\mathbf i_p}\mathbf1\{|Y_{\mathbf i_p}|<\tau_N\}
h_N^{-d_W}K\Bigl(\frac{X_{\mathbf i_p}-w}{h_N}\Bigr)
\Bigm|\Xi_j\Bigr]
\Bigr|\\
&\le
h_N^{d_X}
E\Bigl[h_N^{-d_W}\Bigl|K\Bigl(\frac{X_{\mathbf i_p}-w}{h_N}\Bigr)\Bigr|
\Bigm|\Xi_j\Bigr]\\
&=
h_N^{(1-p)d_X}
\int
\Bigl|K\Bigl(\frac{X_j-w_1}{h_N},\frac{x_{q_2}-w_2}{h_N},\dots,
\frac{x_{q_p}-w_p}{h_N}\Bigr)\Bigr|
\prod_{r=2}^pf(x_{q_r})\,dx_{q_r}\\
&=
\int
\Bigl|K\Bigl(\frac{X_j-w_1}{h_N},s_2,\dots,s_p\Bigr)\Bigr|
\prod_{r=2}^pf(w_r+s_rh_N)\,ds_r\\
&\le C_2B_1^{p-1}.
\end{align*}
The first inequality uses
\(|Y_{\mathbf i_p}|\mathbf1\{|Y_{\mathbf i_p}|<\tau_N\}<\tau_N\) and
\eqref{eq:exchangeable}, by which every ordering contributes the same
conditional expectation up to a permutation of the blocks of \(w\); it
therefore suffices to bound one ordering, in which agent \(j\) occupies
the first position. The second equality conditions on \(X_j\) and
integrates out the covariates of the other \(p-1\) agents, which are
independent of \(\Xi_j\). The third substitutes
\(x_{q_r}=w_r+s_rh_N\), which produces a factor \(h_N^{(p-1)d_X}\). The
last inequality applies Assumption~\ref{aspt:KernelA}\ref{asptpart:KernelA_2} and the density bound
Assumption~\ref{aspt:Model}\ref{asptpart:density}. Taking expectations
of the left side gives the same bound for
\(\tau_N^{-1}h_N^{d_X}|\theta_N(w)|\). Hence
\[
|Q_j(w)|\le2C_2B_1^{p-1}=:b.
\]

\emph{Step 2: Bernstein's inequality.} Write
\[
P\bigl(U_{N,1}(w)>Ma_N\bigr)
=
P\Bigl(\sum_{j=1}^N\tau_N^{-1}h_N^{d_X}\tilde h_{N,1}(\Xi_j;w)
>\frac Mp\,Nh_N^{d_X}a_N\tau_N^{-1}\Bigr)
=P(S\ge t),
\]
where \(S:=\sum_{j=1}^NQ_j(w)\) and
\(t:=(M/p)Nh_N^{d_X}a_N\tau_N^{-1}\). For i.i.d.\ mean-zero summands
bounded by \(b\), Bernstein's inequality gives
\[
P(S\ge t)
\le
\exp\Bigl(-\frac{t^2}{2(v+bt/3)}\Bigr),
\]
with $v:=\sum_{j=1}^NEQ_j(w)^2$. The right side is increasing in \(v\), so for any \(v'\ge v\),
\begin{equation}\label{eq:bernsteinineq}
P(S\ge t)
\le
\exp\Bigl(-\frac{t^2}{2(v'+bt/3)}\Bigr).
\end{equation}

\emph{Step 3: the variance.} Since
\(\tilde h_{N,1}(\Xi_j;w)=\psi_{\{1\}}(\Xi_j;w)-\theta_N(w)\) is a
centered version of the raw projection \(\psi_{\{1\}}\) of
Proposition~\ref{prop:raw-projection-to-Hoeffding} (stated in the
proof of part~\ref{lemmapart:polytailbound2} below),
\(E[\tilde h_{N,1}(\Xi_j;w)^2]\le E[\psi_{\{1\}}(\Xi_j;w)^2]
\le B_2C_1C_2h_N^{-d_X}\) by the second-moment bound proved there with
\(r=1\). Hence
\[
v=\sum_{j=1}^NE\bigl[\tau_N^{-1}h_N^{d_X}\tilde h_{N,1}(\Xi_j;w)\bigr]^2
\le N\tau_N^{-2}h_N^{2d_X}\cdot B_2C_1C_2h_N^{-d_X}
=N\tau_N^{-2}h_N^{d_X}B_2C_2C_1=:v'.
\]

\emph{Step 4: the exponent.} Substituting \(t\), \(v'\), \(b\), and
\(a_N^2=\ln N/(Nh_N^{d_X})\) into \eqref{eq:bernsteinineq},
\begin{align*}
\exp\Bigl(-\frac{t^2}{2(v'+bt/3)}\Bigr)
&=
\exp\left(
-\frac{(\frac Mp Nh_N^{d_X}a_N\tau_N^{-1})^2}
{2\bigl(N\tau_N^{-2}h_N^{d_X}B_2C_2C_1
+2C_2B_1^{p-1}\frac Mp Nh_N^{d_X}a_N\tau_N^{-1}/3\bigr)}
\right)\\
&=
\exp\left(
-\frac{M^2}{2p^2B_2C_2C_1+4pC_2B_1^{p-1}Ma_N\tau_N/3}\,\ln N
\right).
\end{align*}
Since \(\tau_N\ll a_N^{-1}\), \(a_N\tau_N\to0\), so there is
\(N_0\) with \(4pC_2B_1^{p-1}a_N\tau_N/3\le1\) for all \(N>N_0\). For
such \(N\) the exponent is at most
\(-\frac{M^2}{2p^2B_2C_2C_1+M}\ln N\). Choosing \(M=M_\alpha\) large
enough that \(M^2/(2p^2B_2C_2C_1+M)\ge\alpha\) gives
\[
P\bigl(U_{N,1}(w)>M_\alpha a_N\bigr)\le N^{-\alpha}
\qquad\text{for all }N>N_0.
\]

\emph{Step 5: conclusion.} Applying the same argument to
\(-U_{N,1}(w)\), whose summands \(-Q_j(w)\) satisfy the same bounds,
gives the lower tail, and the union of the two gives
\(P(|U_{N,1}(w)|>M_\alpha a_N)\le2N^{-\alpha}\). Neither \(b\), \(v'\),
\(N_0\), nor \(M_\alpha\) depends on \(w\), so the bound holds uniformly:
\[
\sup_{w\in\mathbb R^{d_W}}
P\bigl(|U_{N,1}(w)|>M_\alpha a_N\bigr)
=O(N^{-\alpha}).
\]
\end{proof}

\subsection{Part~\ref{lemmapart:polytailbound2}}

\begin{proof}[Proof of Lemma~\ref{lem:uniformconv},
part~\ref{lemmapart:polytailbound2}]
For \(2\le k\le p\),
\[
U_{N,k}(w)
=
\frac{\binom pk}{\binom Nk}
\sum_{\mathbf q_k\in\mathcal C_{k,N}}
\tilde h_{N,k}(\Xi_{\mathbf q_k};w).
\]
I use a concentration inequality in
Proposition 2.3(c) of \citet{arcones_limit_1993}.

\begin{theorem}[Arcones and Gin\'e]\label{thm:Arcones-and-Gine}
Let \(\zeta_1,\dots,\zeta_N\) be i.i.d.\ with distribution \(P_\zeta\),
and let \(\chi\) be a bounded kernel of order \(k\) that is canonical
for \(P_\zeta\), with \(\|\chi\|_\infty\le c\),
\(E\chi(\zeta_1,\dots,\zeta_k)=0\),
and \(\sigma^2=E\chi(\zeta_1,\dots,\zeta_k)^2\).
There are constants \(c_1,c_2>0\), depending only on \(k\), such that
for all \(t>0\),
\[
P\left(
\left|
N^{-k/2}
\sum_{(i_1,\dots,i_k)\in\mathcal P_{k,N}}
\chi(\zeta_{i_1},\dots,\zeta_{i_k})
\right|
\ge t
\right)\le
c_1\exp\left(
-\frac{c_2t^{2/k}}{
\sigma^{2/k}
+
(ct^{1/k}N^{-1/2})^{2/(k+1)}
}
\right).
\]
\end{theorem}

To apply Theorem~\ref{thm:Arcones-and-Gine}, I first record a proposition that expresses the canonical kernels in terms of the raw projections; it is M\"obius inversion on the lattice of subsets.

\begin{prop}\label{prop:raw-projection-to-Hoeffding}
For \(A\subseteq\{1,\dots,k\}\), define
\(\psi_A(\xi_A;w) := E[\overline Z_{N,\{1,\dots,p\}}(w)\mid\Xi_A=\xi_A]\),
with \(\psi_\varnothing(w):=\theta_N(w)\).
Then
\[
\tilde h_{N,k}(\xi_1,\dots,\xi_k;w)
=
\sum_{A\subseteq\{1,\dots,k\}}
(-1)^{k-|A|}\psi_A(\xi_A;w).
\]
\end{prop}

First, I show that $\tau_N^{-1} h_N^{kd_X} \|\tilde h_{N,k}(\cdot;w)\|_\infty$ is bounded. Let \(r:=|A|\). Conditioning on the \(r\) agents in \(A\) leaves the covariates of
\(p-r\) agents to integrate out. Assumptions~\ref{aspt:Model}\ref{asptpart:density} and
\ref{aspt:KernelA}\ref{asptpart:KernelA_2} imply
\[
\|\psi_A(\cdot;w)\|_\infty
\le
\begin{cases}
\tau_NC_2B_1^{p-r}h_N^{-rd_X},
&1\le r<p,\\
\tau_NK_{\max}h_N^{-pd_X},
&r=p,\\
\tau_NB_1^pC_1,
&r=0.
\end{cases}
\]
Therefore, for all sufficiently large \(N\) such that \(h_N\le1\), by
Proposition~\ref{prop:raw-projection-to-Hoeffding},
\(\|\tilde h_{N,k}(\cdot;w)\|_\infty \le A_k\tau_Nh_N^{-kd_X}\),
where the finite constant
\[
A_k
:=
B_1^pC_1
+
\sum_{r=1}^{k\wedge(p-1)}
\binom kr C_2B_1^{p-r}
+
\mathbf1\{k=p\}K_{\max}
\]
depends only on \(p,k,B_1,C_1,C_2,K_{\max}\).

Second, I show that \(h_N^{kd_X}E[\tilde h_{N,k}(\Xi_1,\dots,\Xi_k;w)^2]\) is bounded. The second-moment bounds for $\psi_A(\Xi_A;w)$ are
\[
E[\psi_A(\Xi_A;w)^2]
\le
\begin{cases}
B_2C_1C_2h_N^{-rd_X},
&1\le r<p,\\
B_2C_1K_{\max}h_N^{-pd_X},
&r=p,\\
B_2B_1^pC_1^2,
&r=0.
\end{cases}
\]
Consider \(1\le r<p\). By the tower property,
\(\psi_A(\Xi_A;w)=E[\widetilde Z_{N,\{1,\dots,p\}}(w)\mid\Xi_A]\), so
\(\psi_A\) integrates out the link shocks
\(\mathcal V_A:=(V_e:e\subseteq A,\ |e|\ge2)\) shared by the agents in
\(A\) (an empty list when \(r=1\)). To bound \(E[\psi_A^2]\) by a cross
moment of the form in Assumption~\ref{aspt:Model}\ref{asptpart:crossmoment},
compare \(\widetilde Z_{N,\{1,\dots,p\}}(w)\) with the truncated kernel
of a second \(p\)-set that shares exactly the agents in \(A\) with
\(\{1,\dots,p\}\), denoted by $\widetilde Z^\dagger:=\widetilde Z_{N,\mathbf q_p^\dagger}(w),$ with $\mathbf q_p^\dagger:=A\cup\{p+1,\dots,2p-r\}.$
By \eqref{eq:polyadic-outcome}, \(\widetilde Z_{N,\{1,\dots,p\}}(w)\) is
a function of \(\Xi_{\{1,\dots,p\}}\) and the shocks \(V_e\),
\(e\subseteq\{1,\dots,p\}\), and \(\widetilde Z^\dagger\) is the same
function of \(\Xi_{\mathbf q_p^\dagger}\) and the shocks \(V_e\),
\(e\subseteq\mathbf q_p^\dagger\). The inputs they have in common are
exactly \((\Xi_A,\mathcal V_A)\). Hence, conditional on \((\Xi_A,\mathcal V_A)\), the two
variables are independent and identically distributed, so
\[
E[\widetilde Z_{N,\{1,\dots,p\}}(w)\mid\Xi_A,\mathcal V_A]^2
=
E[\widetilde Z_{N,\{1,\dots,p\}}(w)\widetilde Z^\dagger\mid\Xi_A,\mathcal V_A].
\]
Since
\(\psi_A(\Xi_A;w)=E\bigl[E[\widetilde Z_{N,\{1,\dots,p\}}(w)\mid\Xi_A,\mathcal V_A]\bigm|\Xi_A\bigr]\),
the conditional Jensen inequality gives
\[
E[\psi_A(\Xi_A;w)^2]
\le
E\bigl[E[\widetilde Z_{N,\{1,\dots,p\}}(w)\mid\Xi_A,\mathcal V_A]^2\bigr]
=
E\bigl[\widetilde Z_{N,\{1,\dots,p\}}(w)\widetilde Z^\dagger\bigr]
\le
E\bigl|\widetilde Z_{N,\{1,\dots,p\}}(w)\widetilde Z^\dagger\bigr|,
\]
with equality in the first step when \(r=1\). The right side is a cross
moment of outcomes on two \(p\)-sets that share exactly \(r\) agents.
Assumption~\ref{aspt:Model} and
the change of variables in the proof of
Theorem~\ref{thm:Variance-bound} bound it by \(B_2C_1C_2h_N^{-rd_X}\).
For \(r=p\), \(\psi_{\{1,\dots,p\}}=\overline Z_{N,\{1,\dots,p\}}(w)\),
and \(E[\overline Z_{N,\{1,\dots,p\}}(w)^2]\le
E[\widetilde Z_{N,\{1,\dots,p\}}(w)^2]\le B_2C_1K_{\max}h_N^{-pd_X}\) by
Jensen's inequality and the diagonal calculation in the proof of
Theorem~\ref{thm:Variance-bound}. For \(r=0\),
\(\theta_N(w)^2\le(E|\widetilde Z_{N,\{1,\dots,p\}}(w)|)^2\le B_2B_1^pC_1^2\)
by the first-moment bound there.

Hence, for all sufficiently large \(N\), by
Proposition~\ref{prop:raw-projection-to-Hoeffding} and the
Cauchy--Schwarz inequality over its \(2^k\) terms,
\(E[\tilde h_{N,k}(\Xi_1,\dots,\Xi_k;w)^2] \le D_kh_N^{-kd_X}\),
where
\[
D_k
:=
2^k\left[
B_2B_1^pC_1^2
+
\sum_{r=1}^{k\wedge(p-1)}
\binom krB_2C_1C_2
+
\mathbf1\{k=p\}B_2C_1K_{\max}
\right].
\]
Consequently,
\[
\sigma_{N,k}^2(w)
:=
E\left[
\left(
\tau_N^{-1}h_N^{kd_X}
\tilde h_{N,k}(\Xi_{\mathbf q_k};w)
\right)^2
\right]
\le D_k\tau_N^{-2}h_N^{kd_X}.
\]

Third, the normalized kernel remains canonical.

Now, I apply Theorem~\ref{thm:Arcones-and-Gine} to $U_{N,k}(w)$
\[
P(|U_{N,k}(w)|\ge Ma_N)
\le
c_1\exp\left(
-\frac{c_2t_{N,k}^{2/k}}{
\sigma_{N,k}^{2/k}
+
(A_kt_{N,k}^{1/k}N^{-1/2})^{2/(k+1)}
}
\right),
\]
where
\[
t_{N,k}
=
Ma_N\frac{\binom Nk}{\binom pk}
N^{-k/2}k!h_N^{kd_X}\tau_N^{-1}.
\]

It suffices to show that
\[
\frac{t_{N,k}^{2/k}}{\sigma_{N,k}^{2/k}\ln N}
\to\infty
\]
and
\[
\frac{t_{N,k}^{2/k}}{
(A_kt_{N,k}^{1/k}N^{-1/2})^{2/(k+1)}
\ln N
}
\to\infty,
\]
since then the exponent in Theorem~\ref{thm:Arcones-and-Gine} exceeds
\(\alpha\ln N\) for all large \(N\), whatever the fixed \(M\).

For the first ratio,
\begin{align*}
\frac{t_{N,k}^{2/k}}{\sigma_{N,k}^{2/k}\ln N}
&\ge
\frac{
\left(
Ma_N
\frac{k!\binom Nk}{\binom pk}
N^{-k/2}
h_N^{kd_X}\tau_N^{-1}
\right)^{2/k}
}{
D_k^{1/k}\tau_N^{-2/k}h_N^{d_X}\ln N
}\\
&=
\frac{
M^{2/k}(k!/\binom pk)^{2/k}
}{
D_k^{1/k}
}
\frac{\binom Nk^{2/k}}N
\frac{a_N^{2/k}h_N^{d_X}}{\ln N}\\
&\ge
\frac{M^{2/k}}{4D_k^{1/k}\binom pk^{2/k}}
a_N^{2/k}\frac{Nh_N^{d_X}}{\ln N}\\
&=
\frac{M^{2/k}}{4D_k^{1/k}\binom pk^{2/k}}
a_N^{2/k-2}\to\infty.
\end{align*}
The second inequality uses the fact that for \(N\ge2k\),
\[
\binom Nk
=
\frac{N^k}{k!}
\prod_{r=0}^{k-1}\left(1-\frac rN\right)
\ge\frac{N^k}{2^kk!},
\]
and the last equality uses \(a_N^2=\ln N/(Nh_N^{d_X})\).
Since \(a_N=o(1)\) and \(k\ge2\), the first ratio diverges.

For the second ratio,
\begin{align*}
&\frac{t_{N,k}^{2/k}}{
(A_kt_{N,k}^{1/k}N^{-1/2})^{2/(k+1)}\ln N
}\\
&\quad=
\left[
\frac{t_{N,k}^2N}{A_k^2(\ln N)^{k+1}}
\right]^{1/(k+1)}\\
&\quad=
\left[
\frac{
M^2(k!/\binom pk)^2
\binom Nk^2N^{-k}
h_N^{(2k-1)d_X}\tau_N^{-2}
}{
A_k^2(\ln N)^k
}
\right]^{1/(k+1)}\\
&\quad\ge
\left[
\frac{M^2}{A_k^24^k\binom pk^2}
\frac{N^kh_N^{(2k-1)d_X}}{(\ln N)^k\tau_N^2}
\right]^{1/(k+1)}
\to\infty.
\end{align*}
The divergence follows from
\(\tau_N\ll(N/\ln N)^{k/2}h_N^{(2k-1)d_X/2}\).
The bound does not depend on \(w\), so for any \(\alpha>0\) and
\(M=M_\alpha\),
\[
\sup_{w\in\mathbb R^{d_W}}P\bigl(|U_{N,k}(w)|>Ma_N\bigr)=O(N^{-\alpha}).
\]
\end{proof}

\subsection{Part~\ref{lemmapart:residual1}}

\begin{proof}[Proof of Lemma~\ref{lem:uniformconv},
part~\ref{lemmapart:residual1}]
Notice that
\begin{align*}
&\left|E[\hat\Psi_N(w)-\widetilde\Psi_N(w)]\right|\\
&\quad\le
E\left[
|Y_{\mathbf q_p}|
\mathbf1\{|Y_{\mathbf q_p}|\ge\tau_N\}
h_N^{-pd_X}
\left|K\!\left(\frac{X_{\mathbf q_p}-w}{h_N}\right)\right|
\right]\\
&\quad\le
\tau_N^{-(s-1)}
E\left[
|Y_{\mathbf q_p}|^s
h_N^{-pd_X}
\left|K\!\left(\frac{X_{\mathbf q_p}-w}{h_N}\right)\right|
\right]\\
&\quad=
\tau_N^{-(s-1)}
\int
E[|Y_{\mathbf q_p}|^s\mid X_{\mathbf q_p}=x_{\mathbf q_p}]
h_N^{-pd_X}
\left|K\!\left(\frac{x_{\mathbf q_p}-w}{h_N}\right)\right|
\prod_{d\in\mathbf q_p}f(x_d)\,dx_d\\
&\quad\le
\tau_N^{-(s-1)}B_{2,s}C_1.
\end{align*}
Since the bound does not depend on \(w\) and
\(\tau_N\gg a_N^{-1/(s-1)}\),
\[
\sup_{w\in\mathbb R^{d_W}}
\left|E[\hat\Psi_N(w)-\widetilde\Psi_N(w)]\right|
\le\tau_N^{-(s-1)}B_{2,s}C_1
=o(a_N).
\]
\end{proof}

\subsection{Part~\ref{lemmapart:residual2}}

\begin{proof}[Proof of Lemma~\ref{lem:uniformconv},
part~\ref{lemmapart:residual2}]
Bounding the kernel by \(K_{\max}\) (Assumption~\ref{aspt:KernelA}\ref{asptpart:KernelA_3}) removes the supremum:
\[
\sup_{w\in\mathbb R^{d_W}}
|\hat\Psi_N(w)-\widetilde\Psi_N(w)|
\le
\frac{K_{\max}}{h_N^{pd_X}M_N}
\sum_{\mathbf i_p\in\mathcal P_{p,N}}
|Y_{\mathbf i_p}|\mathbf1\{|Y_{\mathbf i_p}|\ge\tau_N\}
\le
\frac{K_{\max}\tau_N^{-(s-1)}}{h_N^{pd_X}M_N}
\sum_{\mathbf i_p\in\mathcal P_{p,N}}
|Y_{\mathbf i_p}|^s.
\]
Taking expectations,
\[
E\left[
\sup_{w\in\mathbb R^{d_W}}
|\hat\Psi_N(w)-\widetilde\Psi_N(w)|
\right]
\le
K_{\max}B_3h_N^{-pd_X}\tau_N^{-(s-1)}
=o(a_N),
\]
because
\(\tau_N\gg(a_Nh_N^{pd_X})^{-1/(s-1)}\).
Markov's inequality gives the stated result.
\end{proof}

\subsection{Part~\ref{lemmapart:equalitywpa1}}

\begin{proof}[Proof of Lemma~\ref{lem:uniformconv},
part~\ref{lemmapart:equalitywpa1}]
If every observed outcome is smaller in absolute value than
\(\tau_N\), then the truncated and untruncated statistics agree for
all \(w\). Therefore,
\[
P(\hat\Psi_N=\widetilde\Psi_N)
\ge
P\left(
\max_{\mathbf i_p\in\mathcal P_{p,N}}
|Y_{\mathbf i_p}|<\tau_N
\right).
\]
A union bound and Markov's inequality give
\[
P\Bigl(\max_{\mathbf i_p\in\mathcal P_{p,N}}|Y_{\mathbf i_p}|\ge\tau_N\Bigr)
\le
\sum_{\mathbf i_p\in\mathcal P_{p,N}}P(|Y_{\mathbf i_p}|\ge\tau_N)
\le
M_NB_3\tau_N^{-s}
\le
\frac{B_3}{\phi_N}\to0,
\]
where the last inequality uses \(M_N\le N^p\) and
\(\tau_N^s\ge N^p\phi_N\); the bound is decreasing in \(\tau_N\), so
raising the threshold only helps.
It follows that
\(P(\hat\Psi_N=\widetilde\Psi_N)\to1\).
\end{proof}

\subsection{Part~\ref{lemmapart:linkshock}}

\begin{proof}[Proof of Lemma~\ref{lem:uniformconv},
part~\ref{lemmapart:linkshock}]
Fix \(w\in\mathbb R^{d_W}\). Let \(\Xi:=(\Xi_1,\dots,\Xi_N)\), and let
\(\mathcal E_N\) be the collection of sets \(e\subseteq[N]\) with
\(2\le|e|\le p\). By construction,
the link shocks \((V_e)_{e\in\mathcal E_N}\) are independent of each
other and of \(\Xi\). For \(\mathbf q_p\in\mathcal C_{p,N}\),
\(\widetilde Z_{N,\mathbf q_p}(w)\) depends only on
\(\Xi_{\mathbf q_p}\) and on the shocks \(V_e\) with
\(e\subseteq\mathbf q_p\), so
\(\overline Z_{N,\mathbf q_p}(w)=E[\widetilde Z_{N,\mathbf q_p}(w)\mid\Xi]\).
Hence, conditional on \(\Xi\), \(V_N\) is a function of the independent
shocks.

Let \(\bar K_{\mathbf q_p}:=(p!)^{-1}\sum_{\mathbf i_p\in\mathcal P(\mathbf q_p)}
|K_{\mathbf i_p}(w)|\). Changing one shock \(V_e\) changes only the
summands with \(e\subseteq\mathbf q_p\), and changes each of them by at
most \(2\tau_N\bar K_{\mathbf q_p}\), because the truncated outcomes are
smaller than \(\tau_N\) in absolute value. McDiarmid's inequality
\citep{mcdiarmid_method_1989}, applied conditionally on \(\Xi\), therefore
gives, for every \(t>0\),
\begin{equation}\label{eq:mcdiarmid}
P\bigl(|V_N(w)|\ge t\mid\Xi\bigr)
\le
2\exp\left(-\frac{2t^2}{\sum_{e\in\mathcal E_N}c_e^2}\right),
\qquad
c_e:=\frac{2\tau_N}{\binom Np}\sum_{\mathbf q_p\supseteq e}
\bar K_{\mathbf q_p}.
\end{equation}

Suppose \(K(u)=0\) for \(\|u\|>L_0\).
Write \(w=(w_1',\dots,w_p')'\), let
\(\mathcal B_N:=\bigcup_{a=1}^p\{x\in\mathbb R^{d_X}:\|x-w_a\|\le L_0h_N\}\),
and let \(n_N:=\sum_{i=1}^N\mathbf1\{X_i\in\mathcal B_N\}\).
If \(K_{\mathbf i_p}(w)\ne0\), every agent in \(\mathbf q_p\) has
its covariate in \(\mathcal B_N\). Hence
\(\sum_{\mathbf q_p\supseteq e}\bar K_{\mathbf q_p}\le K_{\max}h_N^{-pd_X}
n_N^{p-|e|}\) if all agents in \(e\) have covariates in
\(\mathcal B_N\), and the sum is zero otherwise. For each \(r\ge2\), at
most \(n_N^r/2\) sets \(e\) with \(|e|=r\) have all their covariates in
\(\mathcal B_N\), so
\[
\sum_{e\in\mathcal E_N}c_e^2
\le
\frac{2\tau_N^2K_{\max}^2}{\binom Np^2}\,
h_N^{-2pd_X}\sum_{r=2}^pn_N^{2p-r}.
\]
The count \(n_N\) is binomial with \(N\) trials and success probability
at most \(pB_1(2L_0h_N)^{d_X}=:c_0h_N^{d_X}\). With
\(\mu_N:=c_0Nh_N^{d_X}\), Bernstein's inequality gives
\(P(n_N>2\mu_N)\le\exp(-3\mu_N/8)\). Since \(a_N=o(1)\),
\(Nh_N^{d_X}/\ln N\to\infty\), so this probability is \(O(N^{-\alpha})\)
uniformly in \(w\). On the event \(\{n_N\le2\mu_N\}\), the bound
\(\binom Np\ge N^p/(2^pp!)\) for \(N\ge2p\) and \(Nh_N^{d_X}\ge1\) give
\[
\sum_{e\in\mathcal E_N}c_e^2
\le
C_V\tau_N^2(Nh_N^{d_X})^{-2},
\qquad
C_V:=2K_{\max}^2(2^pp!)^2\sum_{r=2}^p(2c_0)^{2p-r}.
\]
Taking \(t=Ma_N\) in \eqref{eq:mcdiarmid} and using
\(a_N^2(Nh_N^{d_X})^2=Nh_N^{d_X}\ln N\),
\[
P\bigl(|V_N(w)|\ge Ma_N\bigr)
\le
2\exp\left(-\frac{2M^2}{C_V}\,\frac{Nh_N^{d_X}}{\tau_N^2}\,\ln N\right)
+P(n_N>2\mu_N).
\]
Because \(\tau_N\ll a_N^{-1}\), \(\tau_N^2\le Nh_N^{d_X}\) for all large
\(N\), so the first term is at most \(2N^{-2M^2/C_V}\le2N^{-\alpha}\)
once \(M^2\ge\alpha C_V/2\).
The constants do not depend on \(w\), which proves the claim.
\end{proof}

\section{Proof of
Theorem~\ref{thm:Weak-Uniform-Convergence-for-kernel-estimator}}
\label{app:regression}

The following lemma is used in the proofs of Theorems~\ref{thm:Weak-Uniform-Convergence-for-kernel-estimator} and~\ref{thm:kernel-ratio-clt}.
For a multi-index \(\eta\), write \(\eta!:=\eta_1!\cdots\eta_{d_W}!\)
and \(v^\eta:=v_1^{\eta_1}\cdots v_{d_W}^{\eta_{d_W}}\).

\begin{lemma}[Smoothing bias]\label{lem:bias}
Suppose Assumptions~\ref{aspt:KernelA}\ref{asptpart:KernelA_1}
and~\ref{aspt:kernelB} hold, and let
\(g:\mathbb R^{d_W}\to\mathbb R\) be bounded. Suppose that, for some
\(\delta\in(0,\infty]\), \(g\) belongs to \(\Sigma(\beta,L)\) on
\(\{v:\|v-w\|<\delta\}\) and its derivatives of order at most \(\ell\)
are bounded by \(D\) there. Then
\[
\left|\int K(u)g(w+h_Nu)\,du-g(w)\right|
\le
C_{\mathrm b}h_N^\beta\int\|u\|^\beta|K(u)|\,du,
\]
where
\(C_{\mathrm b}:=c\bigl(L+D\sum_{j=1}^{\ell}\delta^{j-\beta}\bigr)
+2\delta^{-\beta}\sup_v|g(v)|\) and \(c\) depends only on \(\ell\) and
\(d_W\). When \(\delta=\infty\), the terms involving \(\delta\) vanish.
\end{lemma}

\begin{proof}
Because \(\int K(u)\,du=1\), the left side equals
\(|\int K(u)[g(w+h_Nu)-g(w)]\,du|\). Let
\(T(u):=\sum_{1\le|\eta|\le\ell}D^\eta g(w)(h_Nu)^\eta/\eta!\). Since
\(\ell<\beta\), the moments \(\int\|u\|^j|K(u)|\,du\),
\(0\le j\le\ell\), are finite by
Assumptions~\ref{aspt:KernelA}\ref{asptpart:KernelA_1}
and~\ref{aspt:kernelB}, and Assumption~\ref{aspt:kernelB} gives
\(\int K(u)T(u)\,du=0\), so the left side also equals
\(|\int K(u)[g(w+h_Nu)-g(w)-T(u)]\,du|\). If \(\|h_Nu\|<\delta\),
Taylor's expansion with the Lagrange form of the remainder and the
H\"older condition give
\(|g(w+h_Nu)-g(w)-T(u)|\le cL\|h_Nu\|^\beta\). If
\(\|h_Nu\|\ge\delta\), then
\(|g(w+h_Nu)-g(w)|\le2\sup_v|g(v)|(\|h_Nu\|/\delta)^\beta\) and, because
\(j<\beta\),
\(|T(u)|\le cD\sum_{j=1}^\ell\|h_Nu\|^j
\le cD\sum_{j=1}^\ell\delta^{j-\beta}\|h_Nu\|^\beta\). In both cases
the integrand is at most \(C_{\mathrm b}h_N^\beta\|u\|^\beta|K(u)|\).
\end{proof}

\begin{proof}[Proof of Theorem~\ref{thm:Weak-Uniform-Convergence-for-kernel-estimator}]
The proof follows the general approach used in
\citet{hansen_uniform_2008}.
Applying
Theorem~\ref{thm:Weak-Uniform-Convergence} to the numerator and
denominator gives
\[
\begin{aligned}
\sup_{\|w\|\le c_N}
|\hat\Psi_N(w)-E\hat\Psi_N(w)|&=O_p(a_N),\\
\sup_{\|w\|\le c_N}
|\hat f_N(w)-E\hat f_N(w)|&=O_p(a_N).
\end{aligned}
\]
 Both \(\Psi\) and \(f_W\) are bounded: \(|\Psi|\le B_2^{1/2}B_1^{p/2}\)
by the first-moment bound in the proof of
Theorem~\ref{thm:Variance-bound}, and \(f_W\le B_1^p\). Because
\(E\hat\Psi_N(w)=\int K(u)\Psi(w+h_Nu)\,du\) and
\(E\hat f_N(w)=\int K(u)f_W(w+h_Nu)\,du\), applying Lemma~\ref{lem:bias} with
\(\delta=\infty\) gives
\[
\begin{aligned}
\sup_{\|w\|\le c_N}
|E\hat\Psi_N(w)-\Psi(w)|&=O(h_N^\beta),\\
\sup_{\|w\|\le c_N}
|E\hat f_N(w)-f_W(w)|&=O(h_N^\beta).
\end{aligned}
\]
Combining these stochastic and bias bounds,
\[
\begin{aligned}
\sup_{\|w\|\le c_N}
|\hat\Psi_N(w)-\Psi(w)|&=O_p(a_N^*),\\
\sup_{\|w\|\le c_N}
|\hat f_N(w)-f_W(w)|&=O_p(a_N^*).
\end{aligned}
\]
Since \(\Psi=mf_W\),
\[
\hat\Psi_N(w)-m(w)\hat f_N(w)
=
[\hat\Psi_N(w)-\Psi(w)]
-
m(w)[\hat f_N(w)-f_W(w)].
\]
Let
\(E_N:=\{\sup_{\|w\|\le c_N}|\hat f_N(w)-f_W(w)|\le\delta_N/2\}\).
Because \(\delta_N^{-1}a_N^*\to0\), \(P(E_N)\to1\). On \(E_N\),
\(\hat f_N(w)\ge f_W(w)-\delta_N/2\ge\delta_N/2\) for all
\(\|w\|\le c_N\). Hence, on \(E_N\),
\[
\begin{split}
\sup_{\|w\|\le c_N}|\hat m_N(w)-m(w)|
&=
\sup_{\|w\|\le c_N}
\frac{
|\hat\Psi_N(w)-m(w)\hat f_N(w)|
}{
\hat f_N(w)
}\\
&\le
\frac2{\delta_N}
\left(
\sup_{\|w\|\le c_N}|\hat\Psi_N(w)-\Psi(w)|
+
\sup_w|m(w)|
\sup_{\|w\|\le c_N}|\hat f_N(w)-f_W(w)|
\right)\\
&=
O_p(\delta_N^{-1}a_N^*).
\end{split}
\]
Substituting \(h_N\asymp(\ln N/N)^{1/(2\beta+d_X)}\) into \(a_N^*\)
gives the second display.
\end{proof}

\section{Proofs for Section~\ref{sec:pointwise}}
\label{app:pointwise}

Recall \(Z_{N,\mathbf q_p}(w)\) from Appendix~\ref{app:variance}, and
define its analogue for the denominator,
\[
Z^f_{N,\mathbf q_p}(w)
:=
\frac1{p!}
\sum_{\mathbf i_p\in\mathcal P(\mathbf q_p)}K_{\mathbf i_p}(w).
\]
Stack them as
$$A_{N,\mathbf q_p}(w)
:=
\begin{pmatrix}
Z_{N,\mathbf q_p}(w)\\[2pt]
Z^f_{N,\mathbf q_p}(w)
\end{pmatrix}.$$
Let
\(\overline A_{N,\mathbf q_p}(w):=
E[A_{N,\mathbf q_p}(w)\mid\Xi_{\mathbf q_p}]\).

By the tower property and \eqref{eq:exchangeable}, the first projection
defined in Section~\ref{sec:pointwise} can be written as
\[
h_{N,1}(\xi;w)
=
E[\overline A_{N,\{1,\dots,p\}}(w)\mid\Xi_1=\xi]
-
E[\overline A_{N,\{1,\dots,p\}}(w)]
=:
\begin{pmatrix}
h^\Psi_{N,1}(\xi;w)\\[2pt]
h^f_{N,1}(\xi;w)
\end{pmatrix}.
\]

\begin{lemma}[H\'ajek reduction]
\label{lem:pointwise-hajek}
Suppose Assumptions~\ref{aspt:Model}, \ref{aspt:KernelA}, and
\ref{aspt:regularity}\ref{asptpart:regularity1} hold, and
\(Nh_N^{d_X}\to\infty\). At any fixed \(w\),
\[
\sqrt{Nh_N^{d_X}}
\left[
\hat A_N(w)-E\hat A_N(w)
-\frac pN\sum_{i=1}^Nh_{N,1}(\Xi_i;w)
\right]
=o_p(1).
\]
The triangular array
\(\{p\sqrt{h_N^{d_X}/N}\,h_{N,1}(\Xi_i;w):1\le i\le N\}\)
satisfies the Lindeberg condition.
\end{lemma}

\begin{proof}
\emph{H\'ajek reduction.}
The argument is the same for the two coordinates in $\hat A_N(w)$; I give it for the
numerator and note the change for the denominator at the end. Let
\(T_N(w):=\frac pN\sum_{i=1}^Nh^\Psi_{N,1}(\Xi_i;w)\) be the leading
term. By Chebyshev's inequality, it suffices to show
\(\var{\hat\Psi_N(w)-T_N(w)}=o((Nh_N^{d_X})^{-1})\).

Conditioning first on \(\Xi_{\mathbf q_p}\) and then on \(\Xi_i\), and
relabeling the agents by \eqref{eq:exchangeable}, the expression for
\(h_{N,1}\) above gives, for every \(\mathbf q_p\in\mathcal C_{p,N}\)
and every agent \(i\in\mathbf q_p\),
\begin{equation}\label{eq:first-projection}
E[Z_{N,\mathbf q_p}(w)\mid\Xi_i]
=EZ_{N,\mathbf q_p}(w)+h^\Psi_{N,1}(\Xi_i;w).
\end{equation}
Hence
\(\Cov(Z_{N,\mathbf q_p}(w),h^\Psi_{N,1}(\Xi_i;w))=E[h^\Psi_{N,1}(\Xi_1;w)^2]\)
if \(i\in\mathbf q_p\), while \(Z_{N,\mathbf q_p}(w)\) is independent
of \(\Xi_i\) if \(i\notin\mathbf q_p\). Each agent belongs to
\(\binom{N-1}{p-1}=\frac pN\binom Np\) sets in \(\mathcal C_{p,N}\), so
\[
\Cov\bigl(\hat\Psi_N(w),T_N(w)\bigr)
=\frac pN\sum_{i=1}^N\Cov\bigl(\hat\Psi_N(w),h^\Psi_{N,1}(\Xi_i;w)\bigr)
=\frac{p^2}NE[h^\Psi_{N,1}(\Xi_1;w)^2]
=\var{T_N(w)},
\]
and therefore
\(\var{\hat\Psi_N(w)-T_N(w)}=\var{\hat\Psi_N(w)}-\var{T_N(w)}\).

To bound \(\var{\hat\Psi_N(w)}\), recall from the proof of
Theorem~\ref{thm:Variance-bound} that \(\mathcal I_{N,k}\) is the set of
ordered pairs \((\mathbf q_p,\mathbf r_p)\in\mathcal C_{p,N}^2\) that
share exactly \(k\) agents, so that
\[
\var{\hat\Psi_N(w)}
=\binom Np^{-2}\sum_{k=0}^p
\sum_{(\mathbf q_p,\mathbf r_p)\in\mathcal I_{N,k}}
\Cov\bigl(Z_{N,\mathbf q_p}(w),Z_{N,\mathbf r_p}(w)\bigr).
\]
Pairs in \(\mathcal I_{N,0}\) are independent and contribute zero. If
\((\mathbf q_p,\mathbf r_p)\in\mathcal I_{N,1}\) share the agent \(i\),
then \(Z_{N,\mathbf q_p}(w)\) and \(Z_{N,\mathbf r_p}(w)\) share no
other agent primitive and no link shock, so they are independent given
\(\Xi_i\), and \eqref{eq:first-projection} gives
\(\Cov(Z_{N,\mathbf q_p}(w),Z_{N,\mathbf r_p}(w))
=E[h^\Psi_{N,1}(\Xi_1;w)^2]\). Since
\(|\mathcal I_{N,1}|=p\binom Np\binom{N-p}{p-1}\),
\begin{align*}
\var{\hat\Psi_N(w)}-\var{T_N(w)}
&=
\left(\frac{p\binom{N-p}{p-1}}{\binom Np}-\frac{p^2}N\right)
E[h^\Psi_{N,1}(\Xi_1;w)^2]\\
&\quad+
\binom Np^{-2}
\sum_{k=2}^p
\sum_{(\mathbf q_p,\mathbf r_p)\in\mathcal I_{N,k}}
\Cov\bigl(Z_{N,\mathbf q_p}(w),Z_{N,\mathbf r_p}(w)\bigr).
\end{align*}
The first term is nonpositive because
\(\binom{N-p}{p-1}\le\binom{N-1}{p-1}\). The second is at most
\(C\sum_{k=2}^p(Nh_N^{d_X})^{-k}\) by \eqref{eq:Vk-bound},
\eqref{eq:Vp-bound}, and the counting step in the proof of
Theorem~\ref{thm:Variance-bound}. Hence
\[
\var{\sqrt{Nh_N^{d_X}}\,\bigl(\hat\Psi_N(w)-T_N(w)\bigr)}
\le
C\sum_{k=2}^p(Nh_N^{d_X})^{1-k}\to0.
\]
The same argument with \(Y\equiv1\), which satisfies
Assumption~\ref{aspt:Model} with \(B_2=\max\{1,B_1\}^{2p}\), gives the
denominator coordinate, with \(\hat f_N(w)\), \(Z^f_{N,\mathbf q_p}(w)\),
and \(h^f_{N,1}\) in place of \(\hat\Psi_N(w)\), \(Z_{N,\mathbf q_p}(w)\),
and \(h^\Psi_{N,1}\). The higher-order Hoeffding projections and the
link-shock remainder need no separate treatment: both enter the
variance only through the pairs with \(k\ge2\).

\emph{Lindeberg condition.}
Fix an ordered tuple \(\mathbf i_p\) and a position \(a\), and let
\(g_a(\xi):=E[K_{\mathbf i_p}(w)Y_{\mathbf i_p}\mid\Xi_{i_a}=\xi]\).
By the definition of \(h_{N,1}\) and \eqref{eq:exchangeable},
\[
h^\Psi_{N,1}(\xi;w)=\frac1p\sum_{a=1}^p\bigl(g_a(\xi)-Eg_a(\Xi_1)\bigr),
\qquad
E|h^\Psi_{N,1}(\Xi_1;w)|^s\le\frac{2^s}p\sum_{a=1}^pE|g_a(\Xi_1)|^s,
\]
the second by convexity. Write \(\xi=(x,u)\), write \(x_{-a}\) for the
covariates of the agents other than \(i_a\), and let
\(\kappa_x(x_{-a}):=h_N^{-d_W}|K((x_{\mathbf i_p}-w)/h_N)|\prod_{b\ne a}f(x_b)\),
with \(x_{i_a}=x\) in \(x_{\mathbf i_p}\), be the kernel weight times
the density of the other agents' covariates. Conditioning on
\(X_{\mathbf i_p}\) and \(U_{i_a}\) and integrating out \(x_{-a}\),
\[
|g_a(x,u)|
\le
\int
E\bigl[|Y_{\mathbf i_p}|\bigm|X_{\mathbf i_p}=x_{\mathbf i_p},\,U_{i_a}=u\bigr]
\kappa_x(x_{-a})\,dx_{-a}.
\]
The total mass of
\(\kappa_x\) is at most \(\bar\kappa_N:=C_2B_1^{p-1}h_N^{-d_X}\), by
the change of variables in Step~1 of the proof of
Lemma~\ref{lem:uniformconv}\ref{lemmapart:polytailbound1} with
\(A=\{a\}\). Jensen's inequality, applied to the probability measure
\(\kappa_x/\int\kappa_x\) and then to the conditional expectation,
gives
\[
|g_a(x,u)|^s
\le
\bar\kappa_N^{s-1}
\int
E\bigl[|Y_{\mathbf i_p}|^s\bigm|X_{\mathbf i_p}=x_{\mathbf i_p},\,U_{i_a}=u\bigr]
\kappa_x(x_{-a})\,dx_{-a}.
\]
Integrating over \(\Xi_{i_a}=(x,u)\), and using that \(U_{i_a}\) is
independent of the covariates, replaces the conditional moment by
\(E[|Y_{\mathbf i_p}|^s\mid X_{\mathbf i_p}=x_{\mathbf i_p}]\) and
\(\kappa_x\) by the kernel weight times \(\prod_{b=1}^pf(x_b)\); the
resulting integral is bounded by \(B_{2,s}C_1\) exactly as in the
proof of Lemma~\ref{lem:uniformconv}\ref{lemmapart:residual1}. Hence
\(E|g_a(\Xi_1)|^s\le\bar\kappa_N^{s-1}B_{2,s}C_1\) and
\(E|h^\Psi_{N,1}(\Xi_1;w)|^s=O(h_N^{-(s-1)d_X})\); the same argument
with \(Y\equiv1\) gives the bound for \(h^f_{N,1}\). For the scaled
summands,
\[
\sum_{i=1}^N
E\left\|p\sqrt{h_N^{d_X}/N}\;h_{N,1}(\Xi_i;w)\right\|^s
\le
CN^{1-s/2}h_N^{sd_X/2}h_N^{-(s-1)d_X}
=
C(Nh_N^{d_X})^{1-s/2}\to0.
\]
Since \(s>2\), this Lyapunov condition implies the Lindeberg condition.
\end{proof}
\begin{proof}[Proof of Theorem~\ref{thm:kernel-ratio-clt}]
By Lemma~\ref{lem:pointwise-hajek},
\[
\sqrt{Nh_N^{d_X}}
\bigl(\hat A_N(w)-E\hat A_N(w)\bigr)
=
p\sqrt{\frac{h_N^{d_X}}N}
\sum_{i=1}^Nh_{N,1}(\Xi_i;w)
+o_p(1).
\]
For each \(N\), the summands on the right are i.i.d.\ with mean zero,
and their sum has covariance matrix
\(p^2h_N^{d_X}\var{h_{N,1}(\Xi_1;w)}\to\Omega(w)\) by
Assumption~\ref{aspt:pointwise-projection}. They satisfy the Lindeberg
condition by Lemma~\ref{lem:pointwise-hajek}, so the Lindeberg--Feller
central limit theorem and the Cram\'er--Wold device give
\[
\sqrt{Nh_N^{d_X}}
\bigl(\hat A_N(w)-E\hat A_N(w)\bigr)
\to_d N(0,\Omega(w)).
\]

For the bias, choose \(\delta\in(0,1]\) such that the closed ball
\(\{v:\|v-w\|\le\delta\}\) lies in the neighborhood of \(w\) on which
\(f_W\) and \(m\) belong to \(\Sigma(\beta,L)\). Their derivatives of
order at most \(\ell\) are continuous, hence bounded, on this compact
ball, so by the product rule \(\Psi=mf_W\) and \(f_W\) satisfy the
conditions of Lemma~\ref{lem:bias} on \(\{v:\|v-w\|<\delta\}\). Both
functions are bounded on \(\mathbb R^{d_W}\): \(|\Psi|\le
B_2^{1/2}B_1^{p/2}\) by the first-moment bound in the proof of
Theorem~\ref{thm:Variance-bound}, and \(f_W\le B_1^p\). Because
\(E\hat\Psi_N(w)=\int K(u)\Psi(w+h_Nu)\,du\) and
\(E\hat f_N(w)=\int K(u)f_W(w+h_Nu)\,du\), Lemma~\ref{lem:bias} gives
\(E\hat A_N(w)-A(w)=O(h_N^\beta)\). Since \(r\) is continuously
differentiable near \(A(w)\), it follows that
\(m_N^0(w)-m(w)=r(E\hat A_N(w))-r(A(w))=O(h_N^\beta)\).

Applying Theorem~\ref{thm:Variance-bound} to the numerator and, with
\(Y\equiv1\), to the denominator shows that the variance of
\(\hat A_N(w)\) tends to zero; together with \(E\hat A_N(w)\to A(w)\),
this gives \(\hat A_N(w)\to_pA(w)\). Because \(f_W(w)>0\), the event on which
\(\hat f_N(w)\) and \(E\hat f_N(w)\) are both positive has probability
tending to one. On this event, every point on the segment between
\(E\hat A_N(w)\) and \(\hat A_N(w)\) has a positive second coordinate,
so \(r\) is continuously differentiable along the segment and the mean
value theorem gives
\[
\sqrt{Nh_N^{d_X}}
\bigl(\hat m_N(w)-m_N^0(w)\bigr)
=
\nabla r(\bar A_N)'
\sqrt{Nh_N^{d_X}}
\bigl(\hat A_N(w)-E\hat A_N(w)\bigr)
\]
for some \(\bar A_N\) on the segment. Since \(\bar A_N\to_pA(w)\) and
\(\nabla r\) is continuous at \(A(w)\), Slutsky's lemma gives the first
result, with asymptotic variance
\(\nabla r(A(w))'\Omega(w)\nabla r(A(w))=\sigma_m^2(w)\). Under
\(\sqrt{Nh_N^{d_X}}\,h_N^\beta\to0\), the bias \(m_N^0(w)-m(w)\) is
\(o((Nh_N^{d_X})^{-1/2})\), which gives the second result.
\end{proof}


\begin{thebibliography}{28}
\newcommand{\enquote}[1]{``#1''}
\expandafter\ifx\csname natexlab\endcsname\relax\def\natexlab#1{#1}\fi

\bibitem[\protect\citeauthoryear{Aldous}{Aldous}{1981}]{aldous_representations_1981}
\textsc{Aldous, D.~J.} (1981): \enquote{Representations for partially
  exchangeable arrays of random variables,} \emph{Journal of Multivariate
  Analysis}, 11, 581--598.

\bibitem[\protect\citeauthoryear{Arcones and Gin{\'e}}{Arcones and
  Gin{\'e}}{1993}]{arcones_limit_1993}
\textsc{Arcones, M.~A. and E.~Gin{\'e}} (1993): \enquote{Limit theorems for
  {$U$}-processes,} \emph{The Annals of Probability}, 21, 1494--1542.

\bibitem[\protect\citeauthoryear{Aronow, Samii, and Assenova}{Aronow
  et~al.}{2015}]{aronow_cluster-robust_2015}
\textsc{Aronow, P.~M., C.~Samii, and V.~A. Assenova} (2015):
  \enquote{Cluster-robust variance estimation for dyadic data,} \emph{Political
  Analysis}, 23, 564--577.

\bibitem[\protect\citeauthoryear{Bickel, Chen, and Levina}{Bickel
  et~al.}{2011}]{bickel_method_2011}
\textsc{Bickel, P.~J., A.~Chen, and E.~Levina} (2011): \enquote{The method of
  moments and degree distributions for network models,} \emph{The Annals of
  Statistics}, 39, 2280--2301.

\bibitem[\protect\citeauthoryear{Cattaneo, Feng, and Underwood}{Cattaneo
  et~al.}{2024}]{cattaneo_uniform_2024}
\textsc{Cattaneo, M.~D., Y.~Feng, and W.~G. Underwood} (2024): \enquote{Uniform
  inference for kernel density estimators with dyadic data,} \emph{Journal of
  the American Statistical Association}, 119, 2695--2708.

\bibitem[\protect\citeauthoryear{Chen and Christensen}{Chen and
  Christensen}{2015}]{chen_optimal_2015}
\textsc{Chen, X. and T.~M. Christensen} (2015): \enquote{Optimal uniform
  convergence rates and asymptotic normality for series estimators under weak
  dependence and weak conditions,} \emph{Journal of Econometrics}, 188,
  447--465.

\bibitem[\protect\citeauthoryear{Chen and Christensen}{Chen and
  Christensen}{2018}]{chen_optimal_2018}
---\hspace{-.1pt}---\hspace{-.1pt}--- (2018): \enquote{Optimal sup-norm rates
  and uniform inference on nonlinear functionals of nonparametric {IV}
  regression,} \emph{Quantitative Economics}, 9, 39--84.

\bibitem[\protect\citeauthoryear{Chiang, Kato, Ma, and Sasaki}{Chiang
  et~al.}{2022}]{chiang_multiway_2019}
\textsc{Chiang, H.~D., K.~Kato, Y.~Ma, and Y.~Sasaki} (2022): \enquote{Multiway
  cluster robust double/debiased machine learning,} \emph{Journal of Business
  \& Economic Statistics}, 40, 1046--1056.

\bibitem[\protect\citeauthoryear{Chiang and Tan}{Chiang and
  Tan}{2023}]{chiang_empirical_2023}
\textsc{Chiang, H.~D. and B.~Y. Tan} (2023): \enquote{Empirical likelihood and
  uniform convergence rates for dyadic kernel density estimation,}
  \emph{Journal of Business \& Economic Statistics}, 41, 906--914.

\bibitem[\protect\citeauthoryear{Davezies, D'Haultf{\oe}uille, and
  Guyonvarch}{Davezies et~al.}{2021}]{davezies_empirical_2021}
\textsc{Davezies, L., X.~D'Haultf{\oe}uille, and Y.~Guyonvarch} (2021):
  \enquote{Empirical process results for exchangeable arrays,} \emph{The Annals
  of Statistics}, 49, 845--862.

\bibitem[\protect\citeauthoryear{Graham}{Graham}{2020}]{graham_network_2020}
\textsc{Graham, B.~S.} (2020): \enquote{Network data,} in \emph{Handbook of
  Econometrics}, ed. by S.~N. Durlauf, L.~P. Hansen, J.~J. Heckman, and R.~L.
  Matzkin, Amsterdam: North-Holland, vol.~7A, 111--218.

\bibitem[\protect\citeauthoryear{Graham, Niu, and Powell}{Graham
  et~al.}{2021}]{graham_minimax_2021}
\textsc{Graham, B.~S., F.~Niu, and J.~L. Powell} (2021): \enquote{Minimax risk
  and uniform convergence rates for nonparametric dyadic regression,} NBER
  Working Paper 28548, National Bureau of Economic Research,
  \href{https://arxiv.org/abs/2012.08444}{arXiv:2012.08444}.

\bibitem[\protect\citeauthoryear{Graham, Niu, and Powell}{Graham
  et~al.}{2024}]{graham_kernel_2019}
---\hspace{-.1pt}---\hspace{-.1pt}--- (2024): \enquote{Kernel density
  estimation for undirected dyadic data,} \emph{Journal of Econometrics}, 240,
  105336.

\bibitem[\protect\citeauthoryear{Hansen}{Hansen}{2008}]{hansen_uniform_2008}
\textsc{Hansen, B.~E.} (2008): \enquote{Uniform convergence rates for kernel
  estimation with dependent data,} \emph{Econometric Theory}, 24, 726--748.

\bibitem[\protect\citeauthoryear{Jackson}{Jackson}{2008}]{jackson_social_2008}
\textsc{Jackson, M.~O.} (2008): \emph{Social and Economic Networks}, Princeton:
  Princeton University Press.

\bibitem[\protect\citeauthoryear{Kitamura and Laage}{Kitamura and
  Laage}{2024}]{kitamura_estimating_2024}
\textsc{Kitamura, Y. and L.~Laage} (2024): \enquote{Estimating stochastic block
  models in the presence of covariates,} Working paper,
  \href{https://arxiv.org/abs/2402.16322}{arXiv:2402.16322}.

\bibitem[\protect\citeauthoryear{Linton and Nielsen}{Linton and
  Nielsen}{1995}]{linton_kernel_1995}
\textsc{Linton, O. and J.~P. Nielsen} (1995): \enquote{A kernel method of
  estimating structured nonparametric regression based on marginal
  integration,} \emph{Biometrika}, 82, 93--100.

\bibitem[\protect\citeauthoryear{Lov{\'a}sz and Szegedy}{Lov{\'a}sz and
  Szegedy}{2006}]{lovasz_limits_2006}
\textsc{Lov{\'a}sz, L. and B.~Szegedy} (2006): \enquote{Limits of dense graph
  sequences,} \emph{Journal of Combinatorial Theory, Series B}, 96, 933--957.

\bibitem[\protect\citeauthoryear{McDiarmid}{McDiarmid}{1989}]{mcdiarmid_method_1989}
\textsc{McDiarmid, C.} (1989): \enquote{On the method of bounded differences,}
  in \emph{Surveys in Combinatorics, 1989}, ed. by J.~Siemons, Cambridge:
  Cambridge University Press, vol. 141 of \emph{London Mathematical Society
  Lecture Note Series}, 148--188.

\bibitem[\protect\citeauthoryear{Mele}{Mele}{2017}]{mele_structural_2017}
\textsc{Mele, A.} (2017): \enquote{A structural model of dense network
  formation,} \emph{Econometrica}, 85, 825--850.

\bibitem[\protect\citeauthoryear{Menzel}{Menzel}{2021}]{menzel_bootstrap_2021}
\textsc{Menzel, K.} (2021): \enquote{Bootstrap with cluster-dependence in two
  or more dimensions,} \emph{Econometrica}, 89, 2143--2188.

\bibitem[\protect\citeauthoryear{Newey}{Newey}{1994}]{newey_kernel_1994}
\textsc{Newey, W.~K.} (1994): \enquote{Kernel estimation of partial means and a
  general variance estimator,} \emph{Econometric Theory}, 10, 233--253.

\bibitem[\protect\citeauthoryear{Poast}{Poast}{2010}]{poast_misusing_2010}
\textsc{Poast, P.} (2010): \enquote{({M}is)using dyadic data to analyze
  multilateral events,} \emph{Political Analysis}, 18, 403--425.

\bibitem[\protect\citeauthoryear{Sheng}{Sheng}{2020}]{sheng_structural_2020}
\textsc{Sheng, S.} (2020): \enquote{A structural econometric analysis of
  network formation games through subnetworks,} \emph{Econometrica}, 88,
  1829--1858.

\bibitem[\protect\citeauthoryear{Stone}{Stone}{1980}]{stone_optimal_1980}
\textsc{Stone, C.~J.} (1980): \enquote{Optimal rates of convergence for
  nonparametric estimators,} \emph{The Annals of Statistics}, 8, 1348--1360.

\bibitem[\protect\citeauthoryear{Stone}{Stone}{1982}]{stone_optimal_1982}
---\hspace{-.1pt}---\hspace{-.1pt}--- (1982): \enquote{Optimal global rates of
  convergence for nonparametric regression,} \emph{The Annals of Statistics},
  10, 1040--1053.

\bibitem[\protect\citeauthoryear{Stute}{Stute}{1991}]{stute_conditional_1991}
\textsc{Stute, W.} (1991): \enquote{Conditional {$U$}-statistics,} \emph{The
  Annals of Probability}, 19, 812--825.

\bibitem[\protect\citeauthoryear{Tsybakov}{Tsybakov}{2008}]{tsybakov_introduction_2008}
\textsc{Tsybakov, A.~B.} (2008): \emph{Introduction to Nonparametric
  Estimation}, New York: Springer.

\end{thebibliography}
\end{document}